\documentclass[journal]{IEEEtran}
\ifCLASSINFOpdf
\else
\fi
\usepackage{amssymb}
\usepackage{amsmath}
\usepackage{amsthm}

\usepackage{cite}

\usepackage{hyperref}
\usepackage{multirow}
\usepackage{booktabs}

\usepackage{tikz}
\usetikzlibrary{plotmarks}
\usetikzlibrary{calc}
\usetikzlibrary{arrows}
\usetikzlibrary{arrows,intersections}
\usetikzlibrary{arrows.meta}
\usetikzlibrary{shapes,decorations}
\usetikzlibrary{positioning}
\usepackage{pgfplots}
\tikzset{sin v source/.style={
  circle,
  draw,
  append after command={
    \pgfextra{
    \draw
      ($(\tikzlastnode.center)!0.5!(\tikzlastnode.west)$)
       arc[start angle=180,end angle=0,radius=0.425ex]
      (\tikzlastnode.center)
       arc[start angle=180,end angle=360,radius=0.425ex]
      ($(\tikzlastnode.center)!0.5!(\tikzlastnode.east)$)
    ;
    }
  },
  scale=1.5,
 }
}

\pgfplotsset{compat=1.14}

\begin{document}
%
\title{Convex Relaxations of Chance Constrained \\ AC Optimal Power Flow}
%
%

\author{Andreas Venzke,~\IEEEmembership{Student Member, IEEE,}
	Lejla Halilbasic,~\IEEEmembership{Student Member, IEEE,}
	Uros Markovic,~\IEEEmembership{Student Member, IEEE,}
	Gabriela Hug,~\IEEEmembership{Senior Member, IEEE,}
	and Spyros Chatzivasileiadis,~\IEEEmembership{Member, IEEE}
	\thanks{A. Venzke, L. Halilbasic and S. Chatzivasileiadis are with the Department of Electrical Engineering, Technical University of Denmark, Kongens Lyngby, Denmark.}
	\thanks{U. Markovic and G. Hug are with the Power Systems Laboratory, ETH Zurich, Zurich, Switzerland.}
	}

\maketitle

\begin{abstract}
	High penetration of renewable energy sources and the increasing share of stochastic loads require the explicit representation of uncertainty in tools such as the optimal power flow (OPF). Current approaches follow either a linearized approach or an iterative approximation of non-linearities. This paper proposes a semidefinite relaxation of a chance constrained AC-OPF which is able to provide guarantees for global optimality. Using a piecewise affine policy, we can ensure tractability, accurately model large power deviations, and determine suitable corrective control policies for active power, reactive power, and voltage. We state a tractable formulation for two types of uncertainty sets. Using a scenario-based approach and making no prior assumptions about the probability distribution of the forecast errors, we obtain a robust formulation for a rectangular uncertainty set. Alternatively, assuming a Gaussian distribution of the forecast errors, we propose an analytical reformulation of the chance constraints suitable for semidefinite programming. We demonstrate the performance of our approach on the IEEE 24 and 118 bus system using realistic day-ahead forecast data and obtain tight near-global optimality guarantees.
\end{abstract}


\IEEEpeerreviewmaketitle

\section{Introduction}
\IEEEPARstart{P}OWER system operators have to deal with higher degrees of uncertainty in operation and planning. If uncertainty is not explicitly considered, increasing shares of unpredictable renewable generation and stochastic loads, such as electric vehicles, can lead to higher costs and jeopardize system security. The scope of this work is to introduce a convex AC optimal power flow (OPF) formulation which is able to accurately model the effect of forecast errors on the power flow, can define a-priori suitable corrective control policies for active power, reactive power, and voltage, and can provide near-global optimality guarantees.

Chance constraints are included in the OPF formulation to account for uncertainty in power injections, defining a maximum allowable probability of constraint violation. It is generally agreed that the non-linear nature of the AC-OPF along with the probabilistic constraints render the problem for most instances intractable \cite{nemirovski2006convex}. To ensure tractability of these constraints, either a data-driven or scenario-based approach is applied, or the assumption of specific uncertainty distributions is required for an analytical reformulation of the chance constraints. To deal with the higher complexity of chance constrained OPF, existing approaches either assume a DC-OPF \cite{bienstock2014chance,roald2013analytical,lubin2015robust,jabr2015robust,roald2017corrective}, a linearized AC-OPF \cite{guggilam2015scalable, baker2016distribution, summers2014, anese2017chance} or solve iteratively linearized instances of the non-linear AC-OPF \cite{zhang2011chance, schmidli2016stochastic}. Chance constrained DC-OPF results to a faster and more scalable algorithm, but it is an approximation that neglects losses, reactive power, and voltage constraints, and can exhibit substantial approximation errors \cite{Dvijotham2016}.

Refs. \cite{bienstock2014chance} and \cite{roald2013analytical} formulate a chance constrained DC-OPF assuming a Gaussian distribution of the forecast errors. The work in \cite{bienstock2014chance} relies on a cutting-plane algorithm to solve the resulting optimization problem, whereas the work in \cite{roald2013analytical} states a direct analytical reformulation of the same chance constraints. This framework is further extended by the work in \cite{lubin2015robust} which assumes uncertainty sets for both the mean and the variance of the underlying Gaussian distributions to obtain a more distributionally robust formulation. The work in \cite{jabr2015robust} formulates a robust multi-period chance constrained DC-OPF assuming interval bounds on uncertain wind infeeds. These works \cite{bienstock2014chance, roald2013analytical, lubin2015robust, jabr2015robust} include corrective control of the generation units to restore the active power system balance as a function of the forecast errors. The work in \cite{roald2017corrective} extends this corrective control framework to include HVDC converter active power set-points and phase shifting transformers in an N-1 security context.

Alternatively, the works in \cite{guggilam2015scalable,baker2016distribution,summers2014,anese2017chance} use a linearization of the AC power flow equations based on \cite{Dhople2015} to achieve a tractable formulation of the chance constraints. As the operating point is not known a-priori, the linearization is performed around a flat start or no-load voltage, and not the actual operating point.  These works \cite{guggilam2015scalable,baker2016distribution,summers2014,anese2017chance} focus on low-voltage distribution systems with high share of photovoltaic (PV) production and minimize PV curtailment subject to chance constraints on voltage magnitudes. Scenario-based methods are applied to achieve a tractable formulation. In this framework, line flow limits and corrective control from conventional generation are not considered. Furthermore, the utilized linearization in \cite{guggilam2015scalable,baker2016distribution,summers2014,anese2017chance} is designed for radial distribution grids and assumes no voltage control capability of generation units.

In Ref. \cite{zhang2011chance}, an iterative back-mapping and linearization of the full AC power flow equations is used to solve the chance constrained AC-OPF. The recent work in \cite{schmidli2016stochastic} uses an iterative procedure to calculate the full Jacobian, which is the exact AC power flow linearization around the operating point. Assuming a Gaussian distribution of the forecast errors, an analytical reformulation of the chance constraints on voltage magnitude and current line flow is proposed. Although this approach can be shown to scale well, it is not convex and does not guarantee convergence.

In this work, we formulate convex relaxations of chance constrained AC-OPF which allow us to provide guarantees for the optimality of the solution, or otherwise upper-bound the distance to the global optimum of the original non-linear problem. Besides that, we include chance constraints for all relevant state variables, namely active and reactive power, voltage magnitudes and active and apparent branch flows. Two tractable formulations of the chance constraints are proposed. First, based on realistic forecast data and making no prior assumptions about the probability distributions, we formulate a rectangular uncertainty set and, subsequently, the associated chance constraints. Second, assuming a Gaussian distribution of the forecast errors, we provide an analytical reformulation of the chance constraints. 

\subsection{Convex Relaxations and Relaxation Gap}
\begin{figure}
    \begin{footnotesize}
    \tikzstyle{block} = [draw, fill=white, rectangle, 
    minimum height=0em, minimum width=13em]
\tikzstyle{sum} = [draw, fill=blue!20, circle, node distance=4cm]
\tikzstyle{input} = [coordinate]
\tikzstyle{output} = [coordinate]
\tikzstyle{pinstyle} = [pin edge={to-,thin,black}]

\begin{tikzpicture}[node distance=0cm,>=latex']
\node [block] (ACOPF) {$\begin{matrix} \text{(I) Non-convex} \\ \text{AC-OPF} \end{matrix}$};
\node [block, right = 1.5cm of ACOPF] (CCACOPF) {$\begin{matrix} \text{(II) Non-convex chance} \\ \text{ constrained AC-OPF} \end{matrix}$};
\node [block, below = 0.5cm of CCACOPF] (CCACOPFAP) {$\begin{matrix} \text{(III) Non-convex chance} \\ \text{constrained AC-OPF} \\ \text{using affine policy} \end{matrix}$};
\node [block, left = 1.5cm of CCACOPFAP] (CCACOPFAP2) {$\begin{matrix} \text{(IV) Convex chance } \\ \text{constrained AC-OPF} \\  \text{using affine policy} \end{matrix}$};
\node [input, below = 0.25 cm of CCACOPFAP] (tmp){}; 
\draw [->] (ACOPF) -- (CCACOPF);
\draw [->] (CCACOPF) -- (CCACOPFAP);
\draw [->] (CCACOPFAP) --(CCACOPFAP2);
\draw [->] (tmp) -- (CCACOPFAP);
\draw (CCACOPFAP2) |- (tmp);
\node [left = 2.5 cm of tmp] (ss) {};
\node [below = 0.01 cm of ss] (sss) {relaxation gap};
\node [above = 0.8 cm of ss] (sssss) {remove};
\node [above = 0.3 cm of ss] (sssss) {rank-1};
\node [above = 0.07 cm of CCACOPFAP] (sssssss) {\quad parametrize solution space};
\end{tikzpicture}
    \end{footnotesize}
    \caption{We restrict the solution space of the non-linear chance constrained AC-OPF to the parametrization by the affine policy. This problem is relaxed by dropping the non-convex rank constraint. With relaxation gap we refer to the gap between problems (IV) and (III).}
    \label{Relaxation}
\end{figure}
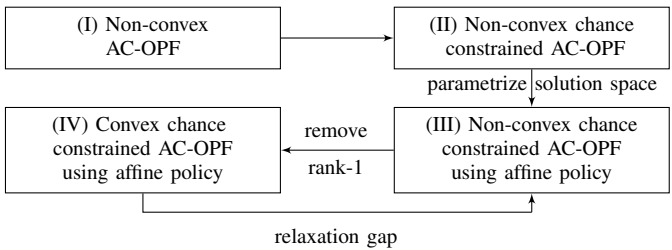
In general, the AC-OPF is a non-convex, non-linear problem. As a result, identified solutions are not guaranteed to be globally optimal and the distance to the global optimum cannot be specified. Recent advancements in the area of convex optimization with polynomials have achieved to relax the non-linear, non-convex optimal power flow problem and transform it to a convex semidefinite (SDP) or second-order cone problem \cite{lavaei2012zero,jabr2007conic,bai2008semidefinite}. Formulating a convex optimization problem results in tractable solution algorithms that can determine the global minimum. Within power systems, finding the global minimum has two important implications. First, from an economic point of view, it can result to substantial cost savings \cite{Cane2012ACOPFOneill}. Second, from  a technical point of view, the global optimum determines a lower or an upper bound of the required control effort. The term relaxation gap denotes the difference between the minimum obtained through the convex relaxation and the global minimum of the original non-convex problem. A relaxation is tight, if the relaxation gap is small. A relaxation is exact, if the relaxation gap is zero, i.e. zero relaxation gap is achieved when the minimum of the convex relaxation coincides with the global minimum of the original non-convex, non-linear problem. Since the work in \cite{Molzahn2011} has shown cases in which the semidefinite relaxation of \cite{lavaei2012zero} fails, it is necessary to investigate the relaxation gap of the obtained solution, and examine the conditions under which we can obtain zero relaxation gap. In the work \cite{madani2015convex} a reactive power penalty is introduced, which allows to upper bound the distance to global optimum. In this work, we develop a penalized semidefinite formulation for a chance constrained AC-OPF, which allows us as well to determine an upper bound of the distance to the global optimum. In Fig.~\ref{Relaxation} we illustrate the previously explained concepts in the context of our work. With relaxation gap, we refer to the gap between the semidefinite relaxation and a non-linear chance constrained AC-OPF which uses the affine policy to parametrize the solution space.
\subsection{Main Contributions}
In this work we propose a framework for a convex chance constrained AC-OPF. The work in \cite{vrakopoulou2013probabilistic} makes a first step towards such a formulation which takes into account security constraints and uncertainty. The change of the system state is described  with an affine policy as an explicit function of the forecast errors. A combination of the scenario approach and robust optimization is used to ensure tractability of the chance constraints \cite{margellos2014road}. For the convex relaxations we build upon the SDP AC-OPF formulation proposed in \cite{lavaei2012zero}. The contributions of our work are the following:
\begin{itemize}
	\item To the best of our knowledge, this is the first paper that proposes a convex formulation for the chance constrained OPF that (a) is able to determine if it has found the global minimum of the original non-convex problem\footnote{As we will discuss later in this paper, in cases where a penalized SDP formulation is necessary, this point corresponds to a near-global minimum.}, and (b) if not, it is able to determine the distance to the global minimum through the relaxation gap. 
    \item In this paper, we introduce a penalty term on power losses which allows us to obtain near-global optimality guarantees and we investigate the conditions under which we can obtain a zero relaxation gap.We show that this penalty term is small in practice, leading to tight near-global optimality guarantees of the obtained solution.
	\item We formulate tractable chance constraints suitable for semidefinite programming for two types of uncertainty sets. First, using a piecewise affine policy, we state a tractable formulation of the chance constrained AC-OPF with convex relaxations that makes no prior assumptions on the type of probability distribution. Using existing data or scenarios, we determine a rectangular uncertainty set; as the set and the chance constraints are affine or convex, we can account for the whole set by enforcing the chance constraints only at its vertices \cite{margellos2014road}. Second, assuming Gaussian distributions, we formulate tractable chance constraints for the optimal power flow equations that are suitable for semidefinite programming. In that, we also assume the correlation of different uncertain variables. To the best of our knowledge, this is the first paper that introduces a tractable reformulation of the chance constrained AC-OPF with convex relaxations for Gaussian distributions.
	\item The proposed framework includes corrective control policies related to active and reactive power, and voltage.
	\item Based on realistic forecast data and the IEEE 118 bus test case, we compare our approach for both uncertainty sets to the chance constrained DC-OPF formulation in \cite{jabr2015robust}, and the iterative AC-OPF in \cite{schmidli2016stochastic}. Compared to the DC-OPF formulation, we find that the formulations proposed in this paper are more accurate and significantly decrease constraint violations. For the rectangular uncertainty set, the affine policy complies with all considered chance constraints and outperforms all other methods having the lowest number of constraint violations. At the same time, we obtain tight near-global optimality guarantees which ensure that the distance to the global optimum is smaller than 0.01\% of the objective value. For a Gaussian distribution, both the iterative AC-OPF and our approach satisfy the constraint violation limit, with our approach achieving slightly lower costs due to the corrective control capabilities. As the realistic forecast data we used do not follow a Gaussian distribution, we also observed that both approaches may exceed the constraint violation limit at certain timesteps for that dataset.
\end{itemize}
\begin{table}[!t]
	\begin{center}
		\caption{Nomenclature} 
		\begin{tabular}{l l}
			\toprule
			\multicolumn{2}{c}{Power grid} \\
						\midrule
			$\mathcal{N}$, $\mathcal{L}$, $\mathcal{G}$ & Set of buses, lines and generators in the power network  \\
			$c_{k_2}$, $c_{k_1}$,  $c_{k_0}$            & Quadratic, linear and constant cost term of generator $k$     \\		$Y$                                         & Admittance matrix                          \\	
			$\bar{y}_{lm}$, $y_{lm}$                    & Shunt and series admittance of line $(l,m)$         \\
			$n_b$                                       & Number of buses in the power network                     \\
			$x_{lm}$                                    & Reactance of line $(l,m)$                               \\
			$B_{\text{AC}}$                             & Admittance matrix based on DC approximation                      \\
						$\text{PTDF}_{lm}$                          & Power transfer distribution factor for line $(l,m)$      \\
			\midrule
			\multicolumn{2}{c}{Optimal power flow} \\
			\midrule
			$P_{G_k}$, $Q_{G_k}$                        & Active and reactive power generation at bus $k$          \\
			$V_k$                                       & Voltage magnitude at bus $k$                             \\
			$P_{lm}$,  $S_{lm}$                         & Active and apparent branch flow on line $(l,m)$          \\
			$\textbf{V}$                                & Complex bus voltage vector                               \\
			$\textbf{X}$                                & Real and imaginary bus voltage vector                    \\
			$P_{D_k}$, $Q_{D_k}$                        & Active and reactive power consumption at bus $k$                \\
			$W$                                         & Matrix with product of voltages                          \\		$W(\zeta_i)$                                & Matrix $W$ as a function of the forecast errors       \\
			$W_0$                                       & Matrix $W$ for forecasted system state                     \\
			$B_i$                                       & System change for forecast error $i$                    \\
			$B_i^u$, $B_i^l$                                      & System change for upper/lower limit on forecast error $i$     \\
			$W_v$                                       & Matrix $W$ for vertex $v$                           \\
						$\rho$                                      & Ratio of second to third eigenvalue of $W$               \\
			$\delta_{\text{opt}}$                       & Near-global optimality measure                           \\
			\midrule
			\multicolumn{2}{c}{Uncertainty modeling} \\
			\midrule
			$n_W$, $\mathcal{W}$                                        & Number of wind farms and set of buses with wind farms                                     \\
			$P_{W}$                                   & Wind infeeds                                 \\
			$P_{W}^f$                                 & Forecasted wind infeeds                        \\
			$\zeta$                                   & Wind forecast errors                            \\
			$\epsilon$                                  & Maximum violation probability of chance constraints      \\
			$d_G$                                       & Generator participation factors                                   \\
			$d_W$                                       & Wind deviation vector                                    \\
			$\gamma$                                  & Slack variable on generator participation factor\\
			$\mu$                                       & Weight for power loss penalty                            \\
			$\cos (\phi)$                               & Power factor of wind farms                               \\
			$\tau$                                      & Ratio of maximum reactive to active power          \\
			$N_{\text{s}}$                              & Number of scenarios                                      \\
			$\beta$                                     & Confidence parameter                                     \\
			$v$                                         & Vertices of rectangular uncertainty set                  \\
			$n_v$, $\mathcal{V}$                               & Number and set of vertices $v$                                     \\
			$\zeta_v$                                   & Forecast error for vertex $v$ of uncertainty set         \\
			$\Lambda$                                   & Covariance matrix                                        \\
			$\lambda$, $\eta$                           & Eigenvalues and eigenvectors of covariance matrix        \\
			$\kappa$                                    & Limit on Gaussian forecast error                         \\
			\bottomrule
		\end{tabular}
		\label{Nomenclature}
	\end{center}
\end{table}
The remainder of this work is structured as follows: In Section~\ref{II} the convex relaxation of the chance constrained AC-OPF problem is formulated. Section~\ref{III} introduces the piecewise affine policy, defines corrective control policies and states the tractable OPF formulation for both uncertainty sets. Section~\ref{IV} states an alternative approach using a linearization based on power transfer distribution factors (PTDFs). Section~\ref{V} investigates the relaxation gap for a IEEE 24 bus system and presents numerical results for a IEEE 118 bus system using realistic forecast data. Section~\ref{VI} concludes the paper. The nomenclature is provided in Table \ref{Nomenclature}. An underline and overline denote, respectively, the upper and lower bound of a variable.

\section{Optimal Power Flow Formulation} \label{II}
\subsection{Convex Relaxation of AC Optimal Power Flow}
For completeness, we outline the semidefinite relaxation of the AC-OPF problem as formulated in \cite{lavaei2012zero}. A power grid consists of $\mathcal{N}$ buses and $\mathcal{L}$ lines. The set of generator buses is denoted with $\mathcal{G}$. The following auxiliary variables are introduced for each bus $k \in \mathcal{N}$ and line $(l,m) \in \mathcal{L}$:
\begin{align}
	Y_k &:= e_k e_k^T Y & \label{aux1} \\
	Y_{lm} &:= (\bar{y}_{lm} + y_{lm}) e_l e_l^T - (y_{lm}) e_l e_m^T  \\
	\textbf{Y}_k & := \dfrac{1}{2} \begin{bmatrix} \Re \{Y_k + Y_k^T\} & \Im \{ Y_k^T - Y_k \} \\ \Im \{ Y_k - Y_k^T\} & \Re \{Y_k + Y_k^T\} \end{bmatrix} & \\
	\textbf{Y}_{lm} &:= \dfrac{1}{2} \begin{bmatrix} \Re \{Y_{lm} + Y_{lm}^T\} & \Im \{ Y_{lm}^T - Y_{lm} \} \\ \Im \{ Y_{lm} - Y_{lm}^T\} & \Re \{Y_{lm} + Y_{lm}^T\} \end{bmatrix} \\
	\bar{\textbf{Y}}_k &:= \dfrac{-1}{2} \begin{bmatrix} \Im \{Y_k + Y_k^T\} & \Re \{ Y_k - Y_k^T \} \\ \Re \{ Y_k^T - Y_k\} & \Im \{Y_k + Y_k^T\} \end{bmatrix} \\
	M_k &:= \begin{bmatrix} e_k e_k^T & 0 \\ 0 & e_k e_k^T \end{bmatrix} \\
	\textbf{X}& : = [\Re \{ \textbf{V} \} \, \Im \{ \textbf{V} \}]^T \label{XXT}
\end{align}
Matrix $Y$ denotes the bus admittance matrix of the power grid, $e_k$ the k-th basis vector, $\bar{y}_{lm}$ the shunt admittance and $y_{lm}$ the series admittance of line $(l,m) \in \mathcal{L}$, and $\textbf{V}$ the vector of complex bus voltages. The non-linear AC-OPF problem can be written using \eqref{aux1} -- \eqref{XXT} as
\begin{align}
	\min_W \sum_{k \in \mathcal{G}} \{ c_{k2} (\text{Tr} \{ \textbf{Y}_k W\} + P_{D_k})^2 + \hphantom{{}=Q_{\text{min},k} } \notag \\[-3\jot]
	c_{k1}  (\text{Tr} \{ \textbf{Y}_k W\} + P_{D_k}) + c_{k0} \} \label{MinGen}
\end{align}
subject to the following constraints for each bus $k \in \mathcal{N}$ and line $(l,m) \in \mathcal{L}$:
\begin{align}
	\underline{P}_{G_k} - P_{D_k} \leq \text{Tr} \{ \textbf{Y}_k W\} \leq \overline{P}_{G_k} - P_{D_k} \label{PBal}                \\
	\underline{Q}_{G_k}  - Q_{D_k} \leq \text{Tr} \{ \bar{\textbf{Y}}_k W\} \leq \overline{Q}_{G_k} - Q_{D_k}\label{QBal}          \\
	\underline{V}_k^2 \leq \text{Tr} \{ M_k W\} \leq \overline{V}_k^2 \label{VCon}                                                 \\
	-\overline{P}_{lm} \leq \text{Tr} \{ \textbf{Y}_{lm} W\} \leq \overline{P}_{lm}  \label{PlmCon}                                \\
	\text{Tr} \{ \textbf{Y}_{lm} W\}^2 +   \text{Tr} \{ \bar{\textbf{Y}}_{lm} W\}^2 \leq \overline{S}_{lm}^2 \label{SlmCon} \\
	W = \textbf{X} \textbf{X}^T \label{WTT}
\end{align}
The objective \eqref{MinGen} minimizes generation cost, where $c_{k2}$, $c_{k1}$ and $c_{k0}$ are quadratic, linear and constant cost variables associated with power production of generator $k \in \mathcal{G}$.\footnote{In case renewable curtailment costs are assumed, this could introduce negative linear costs, which may not result in a tight relaxation.} The terms $P_{D_k}$ and $Q_{D_k}$ denote the active and reactive power consumption at bus $k$. Constraints \eqref{PBal} and \eqref{QBal} include the nodal active and reactive power flow balances; $\underline{P}_{G_k}$, $\overline{P}_{G_k}$, $\underline{Q}_{G_k}$ and  $\overline{Q}_{G_k}$ are generator limits for minimum and maximum active and reactive power, respectively. The bus voltages are constrained by \eqref{VCon} with corresponding lower and upper limits $\underline{V}_k$, $\overline{V}_k$. The active and apparent power branch flow $P_{lm}$ and $S_{lm}$ on line $(l,m) \in \mathcal{L}$ are limited by $\overline{P}_{lm}$ \eqref{PlmCon} and $\overline{S}_{lm}$ \eqref{SlmCon}, respectively. To obtain an optimization problem linear in $W$, the objective function is reformulated using Schur's complement:
\begin{align}
	\min_{W,\, \alpha} \,\, \sum_{k \in \mathcal{G}} \alpha_k \hphantom{100000000000000000000000} \\
	\begin{bmatrix}
	c_{k1} \text{Tr} \{ \textbf{Y}_k W\} + a_k        & \sqrt{c_{k2}} \text{Tr} \{ \textbf{Y}_k W\} + b_k \\
	\sqrt{c_{k2}} \text{Tr} \{ \textbf{Y}_k W\} + b_k & -1                                                \\
	\end{bmatrix} \preceq 0 \label{alpha}
\end{align}
where $a_k : = - \alpha_k + c_{k0} + c_{k1}P_{D_k}$ and $b_k : = \sqrt{c_{k2}} P_{D_k}$. In addition, the apparent branch flow constraint \eqref{SlmCon} is rewritten:
\begin{align}
	\begin{bmatrix}
	- (\overline{S}_{lm})^2                & \text{Tr} \{ \textbf{Y}_{lm} W\} & \text{Tr} \{ \bar{\textbf{Y}}_{lm} W\} \\
	\text{Tr} \{ \textbf{Y}_{lm} W\}       & -1                               & 0                                      \\
	\text{Tr} \{ \bar{\textbf{Y}}_{lm} W\} & 0                                & -1
	\end{bmatrix} \preceq 0
	\label{SlmConSDP}
\end{align}
The non-convex constraint \eqref{WTT} can be expressed by:
\begin{align}
	W \succeq 0 \label{SDP}         \\
	\text{rank}(W) = 1 \label{Rank}
\end{align}
The convex relaxation is introduced by dropping the rank constraint \eqref{Rank}, relaxing the non-linear, non-convex AC-OPF to a convex semidefinite program (SDP). The work in \cite{lavaei2012zero} proves that if the rank of $W$ obtained from the SDP relaxation is 1, then $W$ is the global optimum of the non-linear, non-convex AC-OPF and the optimal voltage vector can be computed following the procedure described in \cite{molzahn2013implementation}.
\subsection{Inclusion of Chance Constraints}
Renewable energy sources and stochastic loads introduce uncertainty in power system operation. To account for uncertainty in bus power injections, we extend the presented OPF formulation with chance constraints. A number of $n_W$ wind farms are introduced in the power grid at buses $k \in \mathcal{W}$ and modeled as
\begin{equation}
	P_{W_k} = P_{W_k}^f + \zeta_k
\end{equation}
where $P_{W}$ are the actual wind infeeds, $P_{W}^f$ are the forecasted values and $\zeta$ are the uncertain forecast errors. To simplify notation, the resulting upper and lower bounds on net active and reactive power injections are written in compact form as:
\begin{align}
	\overline{P}_k & := \overline{P}_{G_k} - P_{D_k} + P^f_{W_k} + \zeta_k, \\
	\underline{P}_k & := \underline{P}_{G_k} - P_{D_k} + P^f_{W_k} +\zeta_k \\
	\overline{Q}_k & := \overline{Q}_{G_k} - Q_{D_k} \label{QLIMUP}\\
	\underline{Q}_k   & := \underline{Q}_{G_k} - Q_{D_k} \label{QLIMDOWN}
\end{align}
The convex chance constrained AC-OPF problem includes chance constraints for each bus $k \in \mathcal{N}$ and line $(l,m) \in \mathcal{L}$:
\begin{align}
	& \hphantom{1000} \min_{W,\, \alpha} \,\, \sum_{k  \in G} \alpha_k \label{ObjCh}                                                            \\
	\text{s.t.\,}     & \text{\eqref{PBal}, \eqref{QBal}, \eqref{VCon}, \eqref{PlmCon}, \eqref{SlmConSDP},  \eqref{alpha}, \eqref{SDP} for } W = W_0 \label{AllCon}     \\
	\mathbb{P} \Big\{ & \underline{P}_k \leq \text{Tr}\{ \textbf{Y}_k W(\zeta) \} \leq \overline{P}_k,  \label{PBalCh}        \\
	& \underline{Q}_k  \leq \text{Tr}\{ \bar{\textbf{Y}}_{k} W(\zeta) \} \leq \overline{Q}_k  \label{QBalCh}  ,  \\
	& \underline{V}_k^2 \leq  \text{Tr}\{ M_{k} W(\zeta)\} \leq \overline{V}_k^2  , \label{VConCh}               \\
	& - \overline{P}_{lm} \leq \text{Tr}\{ \textbf{Y}_{lm} W(\zeta)\} \leq \overline{P}_{lm} \label{PlmConCh} ,\\
	&   \Bigg[ \begin{smallmatrix}
	- (\overline{S}_{lm})^2                        & \text{Tr} \{ \textbf{Y}_{lm} W(\zeta) \} & \text{Tr} \{ \bar{\textbf{Y}}_{lm} W(\zeta) \} \\
	\text{Tr} \{ \textbf{Y}_{lm} W(\zeta) \}       & -1                                       & 0                                              \\
	\text{Tr} \{ \bar{\textbf{Y}}_{lm} W(\zeta) \} & 0                                        & -1
	\end{smallmatrix} \Bigg] \preceq 0 ,
	\label{SlmConCh} \\
	& W(\zeta) \succeq 0 \, \Big\} \geq 1-\epsilon \label{SDPCh}
\end{align}
The parameter $\epsilon \in (0,1)$ defines the upper bound on the violation probability of the chance constraints \eqref{PBalCh} -- \eqref{SDPCh}. The function $W(\zeta)$ denotes the system state as a function of the forecast errors.  The chance constrained AC-OPF problem \eqref{ObjCh} -- \eqref{SDPCh} is an infinite-dimensional problem optimizing over $W(\zeta)$ which is a function of a continuous uncertain variable $\zeta$ \cite{vrakopoulou2013probabilistic}. This renders the problem intractable, which makes it necessary to identify a suitable approximation for $W(\zeta)$ \cite{ben2008selected}.  In the following, an approximation of an explicit dependence of $W(\zeta)$ on the forecast errors is presented.
\section{Piecewise Affine Policy} \label{III}
We present a formulation of the chance constraints using a piecewise affine policy, which approximates the system change as a linear function of the forecast errors. This allows us to include corrective control policies for active and reactive power, and voltages. We propose a tractable formulation for two types of uncertainty sets. First, using an approach based on randomized and robust optimization, and making no prior assumption on the underlying probability distributions, we determine a rectangular uncertainty set. For that, it is sufficient to enforce the chance constraints at its vertices. Second, assuming a Gaussian distribution of the forecast errors, we can provide an analytical reformulation of the linear chance constraints and a suitable approximation of the semidefinite chance constraints.
\subsection{Formulation of Chance Constraints}
The main idea is to describe the matrix $W(\zeta)$ as the sum of the forecasted system operating state $W_0$ and the change of the system state $B_i$ due to each forecast error. Similar to \cite{vrakopoulou2013probabilistic}, the matrix $W(\zeta)$ is approximated using the affine policy
\begin{equation}
	W(\zeta) = W_0 + \sum_{i=1}^{n_\text{w}} \zeta_i B_i \label{AffPol}
\end{equation}
where $W_0$ and $B_i$ are matrices modeled as decision variables. Eq. \eqref{AffPol} provides an affine parametrization of the solution space for the product of real and imaginary part of bus voltages described by $W(\zeta)$. The main advantages of the affine policy are that it resembles affine corrective control policies and naturally allows to include these as well. Furthermore, as the system change depends linearly on the forecast error, in case a Gaussian distribution is assumed, an analytical reformulation can be applied as we will show in Section~\ref{III-E}. Inserting \eqref{AffPol} in \eqref{PBalCh} -- \eqref{SDPCh} yields:
\begin{align}
	\mathbb{P} \Big\{ & \underline{P}_k  \leq \text{Tr}\{ \textbf{Y}_k W_0 \} +  \sum_i^{n_w} \zeta_i \text{Tr}\{ \textbf{Y}_k B_i \}
	\leq \overline{P}_k \label{PBalCh_AP} \\[-1\jot]
	&  \underline{Q}_k \leq \text{Tr}\{ \bar{\textbf{Y}}_{k} W_0 \} + \sum_i^{n_w} \zeta_i \text{Tr}\{ \bar{\textbf{Y}}_{k} B_i \}   \leq \overline{Q}_k  \label{QBalCh_AP} \\[-1\jot]
	&  \underline{V}^2_k \leq  \text{Tr}\{ M_{k} W_0\} +
	\sum_i^{n_w} \zeta_i \text{Tr}\{ M_{k} B_i \} \leq \overline{V}^2_k  \label{VConCh_AP} \\[-2\jot]
	& - \overline{P}_{lm} \leq \text{Tr}\{ \textbf{Y}_{lm} W_0\} +\sum_i^{n_w} \zeta_i \text{Tr}\{ \textbf{Y}_{lm} B_i \} \leq \overline{P}_{lm} \label{PlmConCh_AP}                                                             \\
	&   \begin{bmatrix}
	- \overline{S}_{lm}^2 & \Xi_{lm}^P & \Xi_{lm}^Q \\
	\Xi_{lm}^P            & -1         & 0          \\
	\Xi_{lm}^Q            & 0          & -1
	\end{bmatrix}  \preceq 0 \label{SlmConCH_AP}\\
	&   W_0 + \sum_i^{n_w} \zeta_i B_i \succeq 0 \,  \Big\} \geq 1-\epsilon  \label{SDPCh_AP} 
\end{align}
The terms $\Xi_{lm}^P: =\text{Tr}\{ \textbf{Y}_{lm} W_0\} + \sum_{i=1}^{n_W} \zeta_i \text{Tr}\{ \textbf{Y}_{lm} B_i \}$ and $\Xi_{lm}^Q: =\text{Tr}\{ \bar{\textbf{Y}}_{lm} W_0\}
	+ \sum_{i=1}^{n_W} \zeta_i \text{Tr}\{ \bar{\textbf{Y}}_{lm} B_i \}$  denote the active and reactive power flow on transmission line $(l,m) \in \mathcal{L}$ as a function of the forecast errors. Note that the chance constraints \eqref{PBalCh_AP} -- \eqref{SDPCh_AP} are convex and can be classified in two groups: The constraints \eqref{PBalCh_AP} -- \eqref{PlmConCh_AP} are linear scalar chance constraints and the constraints \eqref{SlmConCH_AP} -- \eqref{SDPCh_AP} are semidefinite chance constraints.
\subsection{Corrective Control Policies}
The affine policy allows to include corrective control policies related to active power, reactive power, and voltage in the AC-OPF formulation. In this work, the implemented policies are generator active power control, generator voltage control, and wind farm reactive power control.

Throughout the transmission system operation, generation has to match demand and system losses. If an imbalance occurs, 
automatic generation control (AGC) restores the system balance \cite{Ibraheem2005}. Hence, designated generators in the power grid will respond to changes in wind power by adjusting their output as part of secondary frequency control. The generator participation factors are defined in the vector $d_G \in \mathbb{R}^{n_b}$. The term $n_b$ denotes the number of buses. The sum of the change in generator active power set-points should compensate the deviation in wind generation, i.\,e.\,$\sum_{k \in G} d_{G_k} = 1$. The wind vector $d^i_{W}\in \mathbb{R}^{n_b}$ for each wind feed-in $i$ in $[1,n_W]$ has a $\{-1\}$ entry corresponding to the bus where the $i$-th wind farm is located at. The other entries are zero. The line losses of the AC power grid vary non-linearly with changes in wind infeeds. To compensate for this change in system losses, we add a slack variable $\gamma_i$ to the generator set-points. This results in the following constraints on each matrix $B_i$, bus $k \in \mathcal{N}$ and wind feed-in $i$ in $[1,n_W]$:
\begin{align}
	\text{Tr} \{ \textbf{Y}_k B_i \}  = d_{G_k} (1 + \gamma_i) + d^i_{W_k} \label{droop}
\end{align}
As a result of \eqref{droop}, it is ensured that each generator compensates the non-linear change in system losses according to its participation factor. To constrain the magnitude of the slack variable, a penalty term is added to the objective function \eqref{ObjCh}, where the term $\mu \geq 0$ is a penalty weight:
\begin{align}
	\min_{W,\, \alpha,\, \gamma} \,\, \sum_{k \in G} \alpha_k + \mu \sum_i^{n_w} \gamma_i \label{ObjPen}
\end{align}
This penalty guides the optimization to a physically meaningful solution, i.e. it allows us to obtain \mbox{rank-1} solution matrices. The increase in losses due to deviations in wind infeeds is minimized. With this penalized semidefinite AC-OPF formulation, near-global optimality guarantees can be derived specifying the maximum distance to the global optimum \cite{madani2015convex}. The numerical results show that while this penalty is necessary to obtain zero relaxation gap, in practice the deviation from the global optimum is very small. This is investigated in detail in Section~\ref{V}.

In power systems, automatic voltage regulators (AVR) are installed as part of the control unit of generators. They keep the voltages at the generator terminals to a value fixed by the operator or a higher level controller \cite{vu1996improved}. The voltage set-point at each generator $k \in \mathcal{G}$ is changed as a function of the forecast errors \cite{vrakopoulou2013probabilistic} and can be retrieved using:
\begin{align}
	V_{k} (\zeta)^2 = & \text{Tr} \{ M_k W_0 \} + \sum_{i=1}^{n_\text{w}} \zeta_i \text{Tr} \{ M_k B_i \}
\end{align}

According to recent revisions in Grid Codes \cite{tsili2009review}, renewable generators such as wind farms have to be able to provide or absorb reactive power up to a certain extent. This is often specified in terms of a power factor $\cos \phi: = \sqrt{\tfrac{P^2}{P^2 + Q^2}}$. In this paper, we include the reactive power capabilities of the wind farms in the optimization. Note that these vary depending on the magnitude of the actual wind infeed. For each $k \in \mathcal{W}$ the constraints \eqref{QLIMUP} and \eqref{QLIMDOWN} are replaced by:
\begin{align}
	\overline{Q}_k & := \overline{Q}_{G_k} - Q_{D_k} + \tau  (P^f_{W_k}+\zeta_k) \label{RPWF1}\\
	\underline{Q}_k   & := \underline{Q}_{G_k} - Q_{D_k} - \tau  (P^f_{W_k}+\zeta_k) \label{RPWF}
\end{align}
where $\tau:=\sqrt{\tfrac{1 - \cos^2\phi}{\cos^2\phi}}$. Using this procedure, active and reactive power set-points of FACTS devices and HVDC converter can also be included in the optimization.
\subsection{Piecewise Affine Policy}\label{III-C}
\begin{figure}
	\center
	\begin{tikzpicture}[
		font = \footnotesize,
		thick,
		>=stealth',
		dot/.style = {
			draw,
			fill = gray!40,
			circle,
			inner sep = 2pt,
			minimum size = 0pt
		},
		dott/.style = {
			draw,
			fill = gray!40,
			star,
			inner sep = 1.5pt,
			minimum size = 0pt,
			star points=4
		}
		]
		\coordinate (O) at (0,0);
		\draw[->] (0,0) -- (8,0) coordinate[label = {below:$P_{W_i}$}] (xmax);
		\draw[->] (0,0) -- (0,4) coordinate[label = {right:System change}] (ymax);
		\draw[dashed] (1.5,1) -- (4.5,1.5);
		\draw[dashed] (1.5,1) -- (6.5,4);
		\draw[dashed] (4.5,1.5) -- (6.5,4);
		\draw[red] plot[smooth] coordinates {(1.5,1) (4.5,1.5) (6.5,4)};
		\path[name path=d] (1.5,1) -- (4.5,1.5) node[dott, label = {below:$W_0$}] {};
		\path[name path=d] (1.5,1) -- (4.5,2.75) node[dott, label = {above:$W_0^\prime$}] {};
		\path[name path=d] (1.5,1) -- (1.5,1) node[dot] {};
		\path[name path=d] (4,1.5) -- (6.5,4) node[dot] {};
		\path[name path=d] (1.5,1) -- (3.5,2.25) node[label = {left:$B_i$}] {};
		\path[name path=d] (4,1.5) -- (5.1,2.32) node[label = {left:$B_i^{u}$}] {};
		\path[name path=d] (1.5,1) -- (3.3,1.15) node[label = {above:$B_i^{l}$}] {};
		\path[name path=d] (4,1.5) -- (7.5,4.2) node[label=below:$W_0 + \overline{\zeta}_i B_i^u$] {};
		\path[name path=d] (4,1.5) -- (7.5,3.8) node[label=above:$W_0^\prime + \overline{\zeta}_i B_i$] {};

		\path[name path=d] (4,1.5) -- (1.0,1.1) node[label=below:$W_0 + \underline{\zeta}_i B_i^l$] {};
		\path[name path=d] (4,1.5) -- (1.0,0.8) node[label=above:$W_0^\prime + \underline{\zeta}_i B_i$] {};

		\draw (1.5,0.1) -- (1.5,-0.1) node[label = {below:$P_{W_i}^f + \underline{\zeta}_i$}] {};
		\draw (4.5,0.1) -- (4.5,-0.1) node[label = {below:$P_{W_i}^f$}] {};
		\draw (6.5,0.1) -- (6.5,-0.1) node[label = {below:$P_{W_i}^f + \overline{\zeta}_i$}] {};

	\end{tikzpicture}
	\caption{Piecewise affine policy: The linearization between upper and lower limit is split into two corresponding piecewise linearizations starting from the exact operating point $W_0$. The red line indicates the true system behavior and the dashed lines the approximation which is made with the corresponding affine policy. This modification allows us to obtain the exact \mbox{rank-1} solution $W_0$, not the higher-rank approximation $W_0^\prime$.}
	\label{ModAff}
\end{figure}
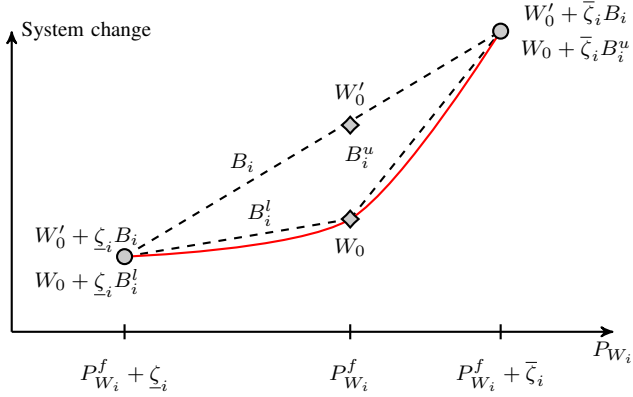
In Fig.\,\ref{ModAff} the affine policy for a wind infeed $P_{W_i}$ is depicted. By choosing an affine policy in the form of \eqref{AffPol}, the maximum and minimum bounds of the uncertainty set are linearly connected using the matrix $B_i$, and the OPF solution at the bounds can be recovered. As the OPF is a non-linear problem, the true system variation will likely not coincide with the linearization. Hence, the affine policy of \cite{vrakopoulou2013probabilistic} is not exact at the operating point $W_0$, but returns only an estimate $W_0^\prime$, i.e. a non-physical higher rank solution. To obtain an exact solution for $W_0$, i.e. a \mbox{rank-1} solution, we introduce a modification to the conventional affine policy by separating the linearization between the maximum and minimum value into an upper part $B_i^u$ and a lower part $B_i^l$, and thereby introducing a piecewise affine policy. Thus, we linearize between the operating point and the maximum and minimum value of the uncertainty set, respectively. We extend the work of \cite{vrakopoulou2013probabilistic}, by ensuring that the obtained solution is exact at the operating point. An additional benefit of our approach is that we get a closer approximation of the true system behavior, while the obtained control policies are piecewise linear.

\subsection{Tractable Formulation for Rectangular Uncertainty Set}
In this section, we provide a tractable formulation of the chance constraints for a rectangular uncertainty set. The proposed procedure is a combination of robust and randomized optimization from \cite{margellos2014road} and which is applied to chance constrained AC-OPF in \cite{vrakopoulou2013probabilistic}. A scenario-based method, which does not make any assumption on the underlying distribution of the forecast errors, is used to compute the bounds of the uncertainty set. Two parameters need to be specified, $\epsilon \in (0,1)$ is the allowable violation probability of the chance constraints and $\beta \in (0,1)$ a confidence parameter. Then, the minimum volume hyper-rectangular set is computed, which with probability $1-\beta$ contains $1-\epsilon$ of the probability mass. According to \cite{vrakopoulou2013probabilistic}, it is necessary to include at least the following number of scenarios $N_{\text{s}}$ to specify the uncertainty set:
\begin{align}
	N_{\text{s}} \geq \dfrac{1}{\epsilon} \dfrac{e}{e-1} (\text{ln} \dfrac{1}{\beta} + 2 n_W - 1) \label{NumberOfSamples}
\end{align}
The term $e$ is Euler's number. The minimum and maximum bounds on the forecast errors $\zeta_i \in [ \underline{\zeta}_i, \overline{\zeta}_i]$ are retrieved by a simple sorting operation among the $N_{\text{s}}$ scenarios and the vertices, i.e. the corner points, of the rectangular uncertainty set can be defined.

To obtain a tractable formulation of the chance constraints, the following result from robust optimization  is used: If the constraint functions are linear, monotone or convex with respect to the uncertain variables, then the system variables will only take the maximum values at the vertices of the uncertainty set \cite{margellos2014road}. The chance constraints \eqref{PBalCh_AP} -- \eqref{PlmConCh_AP} are linear and the semidefinite chance constraints \eqref{SlmConCH_AP}, \eqref{SDPCh_AP} are convex. Hence, it suffices to enforce the chance constraints at the vertices $v \in \mathcal{V}$ of the uncertainty set.

The vector $\zeta_v \in \mathbb{R}^{n_W}$ collects the forecast error bounds for each vertex, i.e. the entries of this vector correspond to the the deviation of each wind farm for a specific vertex $v$. For each vertex, a corresponding slack variable $\gamma_v$ is defined. Based on our experience with the SDP solvers, we introduce the following more numerically robust formulation:
\begin{align}
	W_v: = W_0 + \sum_{i=1}^{n_W} \zeta_{v_i} B_i
\end{align}
The matrix $W_v$ denotes the power flow solution at the corresponding vertex $v$. The active and reactive power limits for each bus $k \in \mathcal{N}$ and vertex $v \in \mathcal{V}$ can be written as:
\begin{align}
	\overline{Q}^v_k  & : =  \overline{Q}_{G_k} - Q_{D_k}  + \tau(P_{W_k}^f + \zeta_{v_k}) \\
	\underline{Q}^v_k & : =  \underline{Q}_{G_k} - Q_{D_k} - \tau( P_{W_k}^f + \zeta_{v_k})  \\
	\overline{P}^v_k & := \overline{P}_{G_k} - P_{D_k} + P^f_{W_k} + \zeta_{v_k}\\
	\underline{P}^v_k & := \underline{P}_{G_k} - P_{D_k} + P^f_{W_k} + \zeta_{v_k}
\end{align}
We provide a tractable formulation of chance constraints \eqref{PBalCh_AP} -- \eqref{SDPCh_AP} for each vertex $v \in \mathcal{V}$, bus $k \in \mathcal{N}$ and line $(l,m) \in \mathcal{L}$:
\begin{align}
	& \underline{P}^v_k  \leq \text{Tr}\{ \textbf{Y}_k W_v \}  \leq \overline{P}^v_k \label{PnodalRect} \\
	&	\underline{Q}^v_k    \leq \text{Tr}\{ \bar{\textbf{Y}}_{k} W_v \}\leq \overline{Q}^v_k    \\
	&	\underline{V}^2_k  \leq  \text{Tr}\{ M_{k} W_v\}   \leq \overline{V}^2_k                               \\
	&	- \overline{P}_{lm} \leq \text{Tr}\{ \textbf{Y}_{lm} W_v\}  \leq \overline{P}_{lm} \\
	&\	\Bigg[ \begin{smallmatrix}
	- (\overline{S}_{lm})^2                   & \text{Tr} \{ \textbf{Y}_{lm} W_v \} & \text{Tr} \{ \bar{\textbf{Y}}_{lm} W_v  \} \\
	\text{Tr} \{ \textbf{Y}_{lm} W_v  \}      & -1                                  & 0                                          \\
	\text{Tr} \{ \bar{\textbf{Y}}_{lm} W_v \} & 0                                   & -1
	\end{smallmatrix} \Bigg] \preceq 0 \label{Slm_Rect}\\
	&	  W_v \succeq 0 \label{SDPRect} \\[-3\jot]
	&	\text{Tr}\{ \textbf{Y}_k (W_v-W_0) \}  =  \sum_{i = 1}^{n_W} \zeta_{v_i} (d_{G_k} (1 + \gamma_v) + d_{W_k}^i) \label{link}
\end{align}
The constraint \eqref{link} links the forecasted system state to each of the vertices. To enforce the semidefinite chance constraint \eqref{SDPCh_AP} for the uncertainty set, it suffices that $W_v$ is positive semidefinite at the vertices of the uncertainty set, i.\,e.\, \eqref{SDPRect} is fulfilled.
\begin{figure}
	\center
	\begin{tikzpicture}[
		font = \footnotesize,
		thick,
		>=stealth',
		dot/.style = {
			draw,
			fill = gray!40,
			circle,
			inner sep = 2pt,
			minimum size = 0pt
		},
		dott/.style = {
			draw,
			fill = gray!40,
			star,
			inner sep = 1.5pt,
			minimum size = 0pt,
			star points=4
		}
		]
		\draw[->] (0,0) -- (6.5,0) coordinate[label = {below:$P_{W_1}$}] (xmax);
		\draw[->] (0,0) -- (0,4.5) coordinate[label = {right:$P_{W_2}$}] (ymax);
		\fill[gray!20] (0.5,0.5)  -- (5.5,0.5) |- (5.5,4.0) -- (0.5,4.0) -- cycle;
		\draw (0.5,0.1) -- (0.5,-0.1) node[label = {below: $P_{W_1}^f + \underline{\zeta}_1 $}] {};
		\draw (2.4,0.1) -- (2.4,-0.1) node[label = {below: $ P_{W_1}^f$}] {};
		\draw (5.5,0.1) -- (5.5,-0.1) node[label = {below: $P_{W_1}^f + \overline{\zeta}_1 $}] {};
		\draw (0.1,0.5) -- (-0.1,0.5);
		\draw (0.1,2.2) -- (-0.1,2.2) node[label = {left: $ P_{W_2}^f$}] {};
		\draw (0.1,4.0) -- (-0.1,4.0);
		\path[name path=d] (0.1,4.0) -- (-0.1,4.25) node[label = {left: $P_{W_2}^f  $}] {};
		\path[name path=d] (0.1,4.0) -- (-0.1,3.75) node[label = {left:$ + \overline{\zeta}_2$}] {};
		\path[name path=d] (0.1,4.0) -- (-0.1,0.75) node[label = {left:$P_{W_2}^f  $}] {};
		\path[name path=d] (0.1,4.0) -- (-0.1,0.25) node[label = {left: $ + \underline{\zeta}_2$}] {};
		\draw[dashed] (2.4,2.2) -- (0.5,0.5);
		\draw[dashed] (2.4,2.2) -- (5.5,0.5);
		\draw[dashed] (2.4,2.2) -- (0.5,4.0);
		\draw[dashed] (2.4,2.2) -- (5.5,4.0);
		\path[name path=d] (1.5,0.0) -- (2.4,2.2) node[dott,label = {below:$W_0$}]{};
		\path[name path=d] (1.5,0.0) -- (0.5,0.5) node[dot,label = {}]{};
		\path[name path=d] (1.5,0.0) -- (5.5,0.5) node[dot,label = {}]{};
		\path[name path=d] (1.5,0.0) -- (0.5,4.0) node[dot,label = {}]{};
		\path[name path=d] (1.5,0.0) -- (5.5,4.0) node[dot,label = {}]{};
		\path[name path=d] (1.5,0.0) -- (0.3,0.25) node[label = {right:$W_3$}]{};
		\path[name path=d] (1.5,0.0) -- (5.7,0.25) node[label = {left:$W_2$}]{};
		\path[name path=d] (1.5,0.0) -- (0.3,4.2) node[label = {right:$W_4$}]{};
		\path[name path=d] (1.5,0.0) -- (5.7,4.2) node[label = {left:$W_1$}]{};
	\end{tikzpicture}
	\caption{Rectangular uncertainty set derived from a scenario-based method displayed for two wind farms. It is sufficient to enforce the chance constraints at the vertices of the uncertainty set. The vertices are denoted with circles.}
	\label{RectSet}
\end{figure}
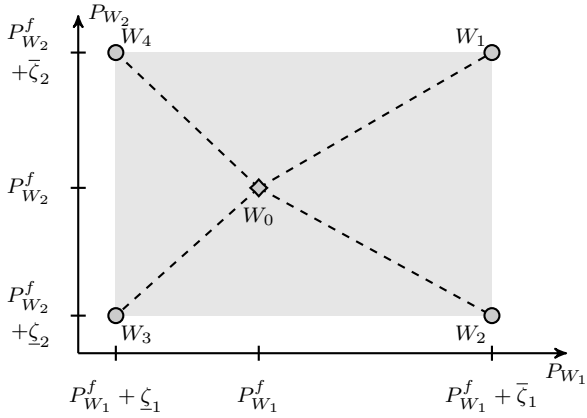
For illustrative purposes, in Fig.\,\ref{RectSet} a rectangular uncertainty set is depicted for two uncertain wind infeeds $P_{W_1}$ and $P_{W_2}$. The resulting optimization problem for a rectangular uncertainty set of dimension $n_w$ minimizes the objective \eqref{ObjPen} subject to constraints \eqref{AllCon} and \eqref{PnodalRect} -- \eqref{link}. Note that the proposed formulation holds for an arbitrary high-dimensional rectangular uncertainty set.
\subsection{Tractable Formulation for Gaussian Uncertainty Set} \label{III-E}
In the following, it is assumed that the forecast errors $\zeta$ are random variables following a Gaussian distribution with zero mean and covariance matrix $\Lambda$. Assuming a Gaussian distribution can be helpful when there is insufficient amount of data at hand, as it can provide a suitable approximation of the power system operation under uncertainty. At the same time, through the covariance matrix, geographical correlations between wind farms, solar PV plants, or other types of uncertainty can be captured. We give a direct tractable formulation of the chance constrained AC-OPF, as the work in \cite{roald2013analytical} presented for the chance constrained DC-OPF.

For a defined confidence interval $1-\epsilon$, the uncertainty set for a Gaussian distribution of the forecast errors is an ellipsoid. First, the direction of linearization of the $B$ matrices is rotated to correspond to the ellipsoid axes which are described by the eigenvectors $\eta_i$ of the covariance matrix. The eigenvalues $\lambda_i$ describe the squared dimension of the ellipsoid in the direction of its axes. Similar to the rectangular uncertainty set, we introduce the following auxiliary variables for each ellipsoid axis $i$ in $[1,n_W]$ and bus $k \in \mathcal{W}$:
	\begin{align}
		\tilde{d}_{G} := d_{G} || \eta_i ||, \, \tilde{d}^i_{W_k} := \eta_i , \, \tilde{\zeta_i} := \sqrt{\lambda_i}
	\end{align}
	With $\tilde{B}_i$ we denote the matrices of the affine policy rotated in the direction of the ellipsoid axes and \eqref{droop} has to hold:
	\begin{align}
		\text{Tr} \{ \textbf{Y}_k \tilde{B}_i \} & = \tilde{d}_{G_k} (1 + \gamma_i) + \tilde{d}^i_{W_k} \label{EQ_Gauss}
	\end{align}
Second, we use theoretical results on chance constraints from the work in \cite{nemirovski2012safe}, which presents the theory for an analytical reformulation of linear scalar chance constraints. To apply the reformulation, we approximate the joint probability of the chance constraint violation \eqref{PBalCh_AP}--\eqref{SDPCh_AP} with the violation probability of each individual chance constraint. Applying the reformulation to the chance constraints \eqref{PBalCh_AP} -- \eqref{PlmConCh_AP} yields for each bus $k \in \mathcal{N}$ and line $(l,m) \in \mathcal{L}$:
\begin{align}
	\underline{P}_k  \leq \text{Tr}\{ \textbf{Y}_k W_0 \}        & \pm
	\sqrt{\sum_i^{n_w} \kappa_i^2 \text{Tr}\{ \textbf{Y}_k \tilde{B}_i \}^2}  \leq \overline{P}_k   \label{PBalSOC} \\
	\underline{Q}_k \leq \text{Tr}\{ \bar{\textbf{Y}}_{k} W_0 \} & \pm
	\sqrt{\sum_i^{n_w} \kappa_i^2 \text{Tr}\{ \bar{\textbf{Y}}_{k} \tilde{B}_i \}^2}  \leq \overline{Q}_k \label{QBalSOC} \\
	\underline{V}^2_k \leq  \text{Tr}\{ M_{k} W_0 \}             & \pm  \sqrt{\sum_i^{n_w} \kappa_i^2 \text{Tr}\{ M_{k} \tilde{B}_i \}^2 }  \leq \overline{V}^2_k                                 \\
	-\overline{P}_{lm} \leq \text{Tr}\{ \textbf{Y}_{lm} W_0 \}   & \pm   \sqrt{\sum_i^{n_w}  \kappa_i^2  \text{Tr}\{ \textbf{Y}_{lm} \tilde{B}_i \}^2 }  \leq \overline{P}_{lm} \label{PlmBalSOC}
\end{align}
The term $\kappa_i:= \Phi^{-1} (1-\epsilon) \tilde{\zeta_i}$ is introduced, where $\Phi^{-1}$ denotes the inverse Gaussian function. The chance constraint \eqref{SDPCh_AP} is a linear matrix inequality which ensures that the matrix $W_0 + \sum_i^{n_w} \tilde{\zeta}_i \tilde{B}_i$  is positive semidefinite inside a confidence interval $1-\epsilon$. An analytical reformulation of this type of constraint is not known \cite{nemirovski2012safe}. As a safe approximation, it suffices to enforce that $W_0 + \sum_i^{n_w} \tilde{\zeta}_i \tilde{B}_i$  is positive semidefinite at maximum corresponding deviations $\pm \kappa_i$ to ensure that \eqref{SDPCh} is fulfilled. We include the following semidefinite constraints for each ellipsoid axis $i \in [1, n_W]$:
\begin{align}
	W_0 \pm \kappa_i \tilde{B}_i \succeq 0 \label{SDPGauss}
\end{align}
This results in \eqref{SDPCh_AP} holding for the outer rectangular approximation of the ellipsoid uncertainty set.
The semidefinite chance constraint on the apparent branch power flow can be conservatively approximated by enforcing it for the smallest rectangular set enclosing the ellipsoid, i.e. by including the constraint \eqref{Slm_Rect} in the optimization.
\begin{figure}
	\center
	\begin{tikzpicture}[
		font = \footnotesize,
		thick,
		>=stealth',
		dot/.style = {
			draw,
			fill = gray!40,
			circle,
			inner sep = 2pt,
			minimum size = 0pt
		},
		dott/.style = {
			draw,
			fill = gray!40,
			star,
			inner sep = 1.5pt,
			minimum size = 0pt,
			star points=4
		}
		]
		\draw[gray!20,fill=gray!20,rotate around={25:(3.25,1.5)}] (3.25,1.5) ellipse (2.5 and 1.75);
		\draw[dotted,rotate around={25:(3.25,1.5)}] (0.75,-0.25) rectangle (5.75,3.25);
		\draw[->] (-0.1,-0.85) -- (6.5,-0.85) coordinate[label = {below:$P_{W_1}$}] (xmax);
		\draw[->] (-0.1,-0.85) -- (-0.1,3.75) coordinate[label = {right:$P_{W_2}$}] (ymax);
		\draw (3.25,-0.75) -- (3.25,-0.95) node[label = {below: $ P_{W_1}^f$}] {};
		\draw (0.0,1.5) -- (-0.2,1.5) node[label = {left: $ P_{W_2}^f$}] {};
		\draw[dashed] (3.25,1.5) --++ (2.2658,1.0565);
		\draw[dashed] (3.25,1.5) --++ (-2.2658,-1.0565);
		\draw[dashed] (3.25,1.5) --++ (-0.7396,1.5860);
		\draw[dashed] (3.25,1.5) --++ (0.7396,-1.5860);

		\path[name path=d,gray!80] (1.5,0.0) -- (4.5,0.9) node[label = {above: (II)}] {};
		\path[name path=d,gray!80] (1.5,0.0) -- (4,3) node[label = {below: (I)}] {};
		\path[name path=d,gray!80] (1.5,0.0) -- (2.5,0.7) node[label = {below: (III)}] {};
		\path[name path=d,gray!80] (1.5,0.0) -- (1.75,1.2) node[label = {above: (IV)}] {};

		\path[name path=d] (1.5,0.0) -- (3.2,1.5) node[label = {below:$W_0$}]{};
		\path[name path=d] (1.5,0.0) -- (3.25,1.5) node[dott,label = {}]{};

		\path[name path=d] (3.25,1.5) --++ (0.7396,-1.5860) node[dot,label = {}]{};
		\path[name path=d] (3.25,1.5) --++ (2.2658,1.0565) node[dot,label = {}]{};
		\path[name path=d] (3.25,1.5) --++ (-2.2658,-1.0565) node[dot,label = {}]{};
		\path[name path=d] (3.25,1.5) --++ (-0.7396,1.5860) node[dot,label = {}]{};

		\path[name path=d] (3.25,1.5) --++ (0.7396,-1.5860) node[label = {right:$W_0 - \kappa_2 \tilde{B}_2^l$}]{};
		\path[name path=d] (3.25,1.5) --++ (2.2658,1.0565) node[label = {right:$W_0 + \kappa_1 \tilde{B}_1^u$}]{};
		\path[name path=d] (3.25,1.5) --++ (-2.2658,-1.0565) node[label = {below:$W_0 - \kappa_1 \tilde{B}_1^l$}]{};
		\path[name path=d] (3.25,1.5) --++ (-0.7396,1.5860) node[label = {left:$W_0 + \kappa_2 \tilde{B}_2^u$}]{};
	\end{tikzpicture}
	\caption{Uncertainty set resulting from a Gaussian distribution of the forecast errors considering correlation. The directions of approximation for the affine policy are rotated corresponding to the eigenvectors of the covariance matrix. The circles denote the points for which the definite chance constraint is enforced. As a result, it holds for the whole dotted rectangular shape. The indices (I) -- (IV) denote the four quadrants of the uncertainty set for each of which the complete set of constraints \eqref{Slm_Rect}, \eqref{EQ_Gauss} and \eqref{PBalSOC} --  \eqref{SDPGauss} is included.}
	\label{GausSet}
\end{figure}
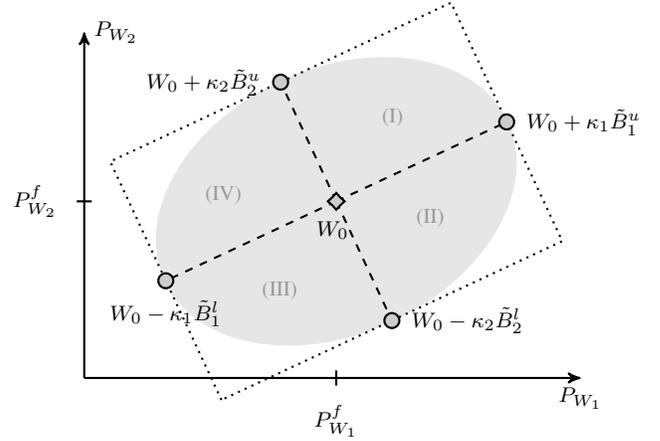

The assumption of a multivariate Gaussian distribution of the forecast errors leads to an uncertainty set which in two dimensions can be described as an ellipse. For the case of two wind farms with uncertain infeeds $P_{W_1}$ and $P_{W_2}$ this configuration is depicted in Fig.\,\ref{GausSet}. Incorporating the results on the modification of the affine policy presented in Section~\ref{III-C}, we add the constraints \eqref{PBalSOC} -- \eqref{PlmBalSOC} not for $\tilde{B}_i$ but for both $\tilde{B}_i^u$ and $\tilde{B}_i^l$ and each of their combinations, splitting the uncertainty set into four quadrants (I) -- (IV) as depicted in Fig.\,\ref{GausSet}. The resulting optimization problem corresponds to minimizing objective function \eqref{ObjPen} subject to constraints \eqref{AllCon}, \eqref{Slm_Rect}, \eqref{EQ_Gauss} and \eqref{PBalSOC} --  \eqref{SDPGauss} for each quadrant of the ellipsoid.

\section{Linearization using PTDFs}\label{IV}
In the following, an alternative approach is presented which is used as benchmark for comparison with the rest of the approaches presented in this paper. To describe the system change as a function of the forecast errors, in this section we introduce a linear approximation based on DC power flow. This linear approximation uses the so-called power transfer distribution factors (PTDFs) to estimate the change in line loading due to a change in active power injections. This approach has been used in the works in \cite{roald2013analytical} and \cite{chatzivasileiadis2011flexible} in the context of DC- and AC-OPF, respectively.

The PTDFs use the DC power flow representation, i.\,e.\, assuming that the voltage magnitudes of all buses are equal to 1 p.u. and the resistances of branches are neglected. Hence, line losses are neglected and the generator participation factors are defined without including the slack term $\gamma$. As we assume constant voltage magnitudes, the semidefinite \eqref{SDPCh}, the voltage \eqref{VConCh} and the reactive power \eqref{QBalCh} chance constraints are dropped and the focus is on approximating the chance constraints for the active power bus injection and active power branch flow, Eqs.~\eqref{PBalCh} and \eqref{PlmConCh}. The admittance matrix $B_{\text{DC}}$ is constructed using only the line reactances $x_{lm}$. The resulting matrix is singular. Thus, one column and the corresponding row are removed to obtain $\tilde{B}_{\text{DC}}$. The vectors $d_G$ and $d^i_W$ collect the generator participation factors and wind injections, and $\tilde{d}_G$ and $\tilde{d}^i_W$ denote the corresponding vectors with the first entry removed. The PTDF for each line $(l,m) \in \mathcal{L}$ is defined as follows:
\begin{align}
	\text{PTDF}_{lm} = (e_l - e_m)^T \tfrac{1}{x_{lm}} \tilde{B}_{\text{DC}}^{-1} \label{LF}
\end{align}
The PTDFs provide an approximate linear relation between a change in bus power injections and the change of the active power flow over a transmission line. Assuming the maximum and minimum bounds of the forecast errors are described by a rectangular uncertainty set with vertices $\zeta_v$ from the previously described scenario-based approach, we formulate a tractable approximation of \eqref{PBalCh} and \eqref{PlmConCh} for each bus $k \in \mathcal{N}$, line $(l,m) \in \mathcal{L}$ and vertex $v \in \mathcal{V}$:
\begin{align}
	\underline{P}_{k}^v & \leq  \text{Tr}\{ \textbf{Y}_k W_0 \} + \sum_i^{n_W} \zeta_{v_i} (d_{G_k}+d^i_{W_k}) \leq \overline{P}_{k}^{v}\\[-2\jot]
	-\overline{P}_{lm}  & \leq \text{Tr}\{ \textbf{Y}_{lm} W_0 \} + \nonumber \\
& \quad \quad 	\quad \quad \quad \sum_i^{n_W} \text{PTDF}_{lm} \zeta_{v_i} (\tilde{d}_{G}+\tilde{d}^i_{W})  \leq \overline{P}_{lm} 
\end{align}
Assuming the forecast errors follow a Gaussian distribution with zero mean and co-variance matrix $\Lambda$, we formulate a tractable approximation of \eqref{PBalCh} and \eqref{PlmConCh} for each bus $k \in \mathcal{N}$ and line $(l,m) \in \mathcal{L}$:
\begin{align}
	\underline{P}_{k}  \leq \text{Tr}\{ \textbf{Y}_k W_0 \} \pm  \Phi^{-1} (1-\epsilon)
	\sqrt{ d_{G_k}^2 \mathbf{1}^T \Lambda \mathbf{1} }\leq \overline{P}_{k} \label{PgenDC}                 \\
	-\overline{P}_{lm}  \leq \text{Tr}\{ \textbf{Y}_{lm} W_0 \} \pm \Phi^{-1} (1-\epsilon) \sqrt{ \Psi^T \Lambda \Psi  }   \leq \overline{P}_{lm}\label{PlmDC}
	 \end{align} 
The term $\mathbf{1}\in \mathbb{R}^{n_W}$ denotes the vectors of ones. The vector $\Psi \in \mathbb{R}^{n_W}$ contains for each wind feedin $i \in [1,n_W]$ the approximated change in line loading: \begin{align}
\Psi_i = \text{PTDF}_{lm} (\tilde{d}_{G}+\tilde{d}^i_{W})
\end{align}

\section{Simulation And Results}\label{V}
In this section, we first describe the simulation setup. Subsequently, using the IEEE 24 bus test case, we investigate the relaxation gap of the obtained solution matrices as a function of the penalty weight. Detailed results on the IEEE 118 bus test case using realistic forecast data are provided and our proposed approaches are compared to two alternative approaches described in the literature.
\subsection{Simulation Setup}
The optimization problem is implemented in Julia using the optimization toolbox JuMP \cite{lubin2015computing} and the SDP solver MOSEK 8 \cite{mosek}. A small resistance of $10^{-4}$ has to be added to each transformer, which is a condition for obtaining zero relaxation gap \cite{lavaei2012zero}. To investigate whether the relaxation gap of an obtained solution matrix $W$ is zero, the ratio $\rho$ of the $2^{\text{nd}}$ to $3^{\text{rd}}$ eigenvalue is computed, a measure proposed by \cite{molzahn2013implementation}. This value should be around $10^5$ or larger for zero relaxation gap to hold, which means that the obtained solution matrix is rank-2. The respective \mbox{rank-1} solution can be retrieved by following the procedure described in \cite{molzahn2013implementation}. According to \cite{lavaei2012zero}, the obtained solution is then a feasible solution to the original non-linear AC-OPF problem.

The work in \cite{madani2015convex} proposes the use of the following measure to evaluate the degree of  the near-global optimality of a penalized SDP relaxation. Let $\tilde{f}_1(x)$ be the generation cost of the convex OPF without a penalty term and $\tilde{f}_2(x)$ the generation cost of the convex OPF with a penalty weight sufficiently high to obtain \mbox{rank-1} solution matrices. Then, the near-global optimality can be assessed by computing the parameter $\delta_\text{opt}:= \tfrac{\tilde{f}_1(x)}{\tilde{f}_2(x)} \cdot 100 \%$. The closer this parameter is to 100\%, the closer the solution is to the global optimum. Note that this distance is an upper bound to the distance from global optimality.

\subsection{Investigating the Relaxation Gap}
\label{RelGap}
This section investigates the relaxation gap of the obtained matrices. With relaxation gap, we refer to the gap between the SDP relaxation and a non-linear chance constrained AC-OPF which uses the affine policy to parametrize the solution space. The IEEE 24 bus system with parameters specified in \cite{zimmerman2011matpower} is used. The allowable violation probability is selected to be $\epsilon=5\%$. Two wind farms with a forecasted infeed of 50 MW and 150 MW and a maximum power of 150 MW and 400 MW are introduced at buses 8 and 24, respectively. For illustrative purposes, the forecast error for the rectangular uncertainty is assumed to be bounded within $\pm 50\%$ of the forecasted value with 95\% probability. For the Gaussian uncertainty set, a standard deviation of $25\%$ of the forecasted value and no correlation between both wind farms is assumed. Each generator adjusts its active power proportional to its maximum active power to react to deviations in wind power output.
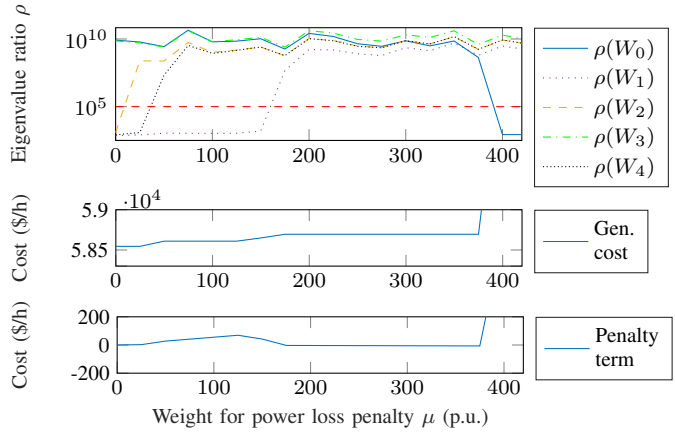
\begin{figure}
		\begin{footnotesize}
%
%
\definecolor{mycolor1}{rgb}{0.00000,0.44700,0.74100}%
\definecolor{mycolor2}{rgb}{0.85000,0.32500,0.09800}%
\definecolor{mycolor3}{rgb}{0.92900,0.69400,0.12500}%
\definecolor{mycolor4}{rgb}{0.49400,0.18400,0.55600}%
\definecolor{mycolor5}{rgb}{0.46600,0.67400,0.18800}%
\begin{tikzpicture}

\begin{axis}[%
width=5.4cm,
height=1.5cm,
at={(0.758in,0.481in)},
scale only axis,
xmin=0,
xmax=420,
xlabel style={font=\color{white!15!black}},
ymode=log,
ymin=100,
ymax=1000000000000,
ytick = {100,100000,100000000000},
yticklabels = {$\quad \quad \quad$,$10^5$,$10^{10}$},
yminorticks=true,
ylabel style={font=\color{white!15!black}},
ylabel={Eigenvalue ratio $\rho$},
axis background/.style={fill=white},
legend pos=outer north east,
legend style={legend cell align=left, cells={align=left}, align=left, draw=white!15!black}
]
\addplot [color=mycolor1,solid]
  table[row sep=crcr]{%
0	78570687754.0553\\
25	53867805216.4181\\
50	20766361456.8135\\
75	603736511706.823\\
100	55576363271.6182\\
125	64282261211.1811\\
150	104795405101.371\\
175	12792731989.9645\\
200	314152489658.106\\
225	167220639861.497\\
250	38067824400.0582\\
275	22324620363.6873\\
300	69047150336.1932\\
325	24442009656.2389\\
350	66553953572.7641\\
375	2304205954.37493\\
400	331.33781510908\\
425	327.817582302869\\
450	331.066667518568\\
475	335.525579644997\\
500	338.657870709993\\
};
\addlegendentry{$\rho(W_0)$}

\addplot [color=mycolor4,dotted]
  table[row sep=crcr]{%
0	270.861878396884\\
25	312.211242428608\\
50	440.472998211468\\
75	430.303787464962\\
100	427.592004111881\\
125	433.43833205083\\
150	670.290122449778\\
175	158314562.746653\\
200	11627948703.6154\\
225	10650109474.815\\
250	4859050395.8258\\
275	3411989447.37241\\
300	16706976155.6495\\
325	9168772458.99105\\
350	43250171504.0315\\
375	3550776714.43578\\
400	23968847202.4977\\
425	11257298268.5113\\
450	1408254454.23901\\
475	16265661161.5399\\
500	9841965418.66314\\
};
\addlegendentry{$\rho(W_1)$}

\addplot [color=mycolor3,dashed]
  table[row sep=crcr]{%
0	503.76312857151\\
25	1149330317.13164\\
50	1106302361.22532\\
75	49337645933.8232\\
100	6578587418.86449\\
125	10386211273.1377\\
150	19194117250.7687\\
175	3309778906.72227\\
200	106838052324.2\\
225	73071263058.5865\\
250	24194473823.0992\\
275	16572078662.5929\\
300	69440264975.391\\
325	35980508672.8314\\
350	159648820312.85\\
375	11932790680.1935\\
400	80789351319.5777\\
425	36316630429.3265\\
450	4491458558.76096\\
475	49632794571.6971\\
500	28532167976.419\\
};
\addlegendentry{$\rho(W_2)$}

\addplot [color=green,dashdotted]
  table[row sep=crcr]{%
0	67442320886.2708\\
25	44848859636.7124\\
50	19254036853.4715\\
75	543493224047.853\\
100	57930024483.0546\\
125	85111982550.2246\\
150	139864487404.985\\
175	19146683324.4835\\
200	545472211348.661\\
225	331206672645.824\\
250	91808533955.0772\\
275	65220695691.9539\\
300	234309008384.783\\
325	137283550017.543\\
350	529983358365.9\\
375	32660010666.7769\\
400	220212663890.612\\
425	90750376147.7821\\
450	9827839471.59982\\
475	103044750113.027\\
500	44492852294.1966\\
};
\addlegendentry{$\rho(W_3)$}

\addplot [color=black,densely dotted]
  table[row sep=crcr]{%
0	313.477392538646\\
25	506.457060969291\\
50	59955991.2579128\\
75	26954641594.1981\\
100	5297597014.86487\\
125	9754953261.57577\\
150	19420102717.2971\\
175	3327611908.78621\\
200	107275953433.796\\
225	69985462322.1644\\
250	22295605247.9848\\
275	16616135774.5185\\
300	71456840738.6051\\
325	36210542501.7854\\
350	162743979542.718\\
375	12079914429.1648\\
400	81749430578.6987\\
425	36543199205.3493\\
450	4544559018.79151\\
475	48937393839.2503\\
500	28500849669.2749\\
};
\addlegendentry{$\rho(W_4)$}
    \addplot [color=red,dashed]
      table[row sep=crcr]{%
      0 100000 \\
      500 100000 \\
      };
\end{axis}
\end{tikzpicture}%
%
%
\definecolor{mycolor1}{rgb}{0.00000,0.44700,0.74100}%
\begin{tikzpicture}

\begin{axis}[%
width=5.4cm,
height=0.75cm,
at={(0.758in,0.481in)},
scale only axis,
xmin=0,
xmax=420,
xlabel style={font=\color{white!15!black}},
ymin=58300,
ymax=59000,
ytick = {58300,58500,59000},
yticklabels = {$\quad \quad \quad$,5.85,5.9},
ylabel style={font=\color{white!15!black}},
ylabel={Cost (\$/h)},
legend pos=outer north east,
axis background/.style={fill=white},
legend style={legend cell align=left,cells={align=left}, align=left, draw=white!15!black}
]
\addplot [color=mycolor1]
  table[row sep=crcr]{%
0	58545.4020245022\\
25	58545.4022225959\\
50	58609.6638409768\\
75	58609.6646826412\\
100	58609.6646191674\\
125	58609.6657116499\\
150	58649.7522619478\\
175	58695.501028542\\
200	58695.5018712662\\
225	58695.5017283517\\
250	58695.5017072166\\
275	58695.5015945566\\
300	58695.5015575103\\
325	58695.5016483018\\
350	58695.501857438\\
375	58695.5017655773\\
400	61167.3106018006\\
425	61167.3096422303\\
450	61167.3114679178\\
475	61167.3120971494\\
500	61167.3113251682\\
};
\addlegendentry{Gen. \\ cost}

\end{axis}
\end{tikzpicture}%
%
%
\definecolor{mycolor1}{rgb}{0.00000,0.44700,0.74100}%
\begin{tikzpicture}

\begin{axis}[%
width=5.4cm,
height=0.75cm,
at={(0.758in,0.481in)},
scale only axis,
xmin=0,
xmax=420,
xlabel style={font=\color{white!15!black}},
xlabel={Weight for droop penalty $\mu$ (p.u.)},
ymin=-200,
ymax=200,
ytick = {-200,0,200},
yticklabels = {-200,$\quad \quad \, \, \,0$,200},
ylabel style={font=\color{white!15!black}},
ylabel={Cost (\$/h)},
axis background/.style={fill=white},
legend pos=outer north east,
legend style={legend cell align=left, cells={align=left}, align=left, draw=white!15!black}
]
\addplot [color=mycolor1]
  table[row sep=crcr]{%
0	-0\\
25	2.51591055033679\\
50	27.6586516977893\\
75	41.4885778348734\\
100	55.3177540335896\\
125	69.1449893691915\\
150	42.1772242840145\\
175	-3.08678948655681\\
200	-3.52804747245641\\
225	-3.96903285554956\\
250	-4.4099401920248\\
275	-4.85094313961587\\
300	-5.28693581441569\\
325	-5.73294210343789\\
350	-6.17403852880823\\
375	-6.61490827805986\\
400	820.212395210869\\
425	871.475757620454\\
450	922.739355611427\\
475	974.002030292799\\
500	1025.26566280394\\
};
\addlegendentry{Penalty\\term}

\end{axis}
\end{tikzpicture}%
		\end{footnotesize}
	\caption{Eigenvalue ratios $\rho$, generation cost and penalty term as a function of the power loss penalty weight $\mu$ for a IEEE 24 bus test case with two wind farms and a rectangular uncertainty set.}
	\label{EV_ratio}
\end{figure}

For the rectangular uncertainty set, Fig.~\ref{EV_ratio} shows the eigenvalue ratios $\rho$ of the matrices $W_0 - W_4$ as a function of the penalty weight $\mu$. A certain minimum value for the weight $\mu = 175$ is necessary to obtain solution matrices with \mbox{rank-1}, i.e. eigenvalue ratio $\rho$ higher than $10^5$, at the operating state $W_0$ and the four vertices of the rectangular uncertainty set $W_1 - W_4$. The near-global optimality at $\mu = 175$ for this test case evaluates to a tight upper bound of $99.74\%$. If the penalty weight is increased beyond $\mu = 375$ a higher rank solution is obtained for the forecasted system state.
\begin{figure}
		\begin{footnotesize}
%
%
\definecolor{mycolor1}{rgb}{0.00000,0.44700,0.74100}%
\definecolor{mycolor2}{rgb}{0.85000,0.32500,0.09800}%
\definecolor{mycolor3}{rgb}{0.92900,0.69400,0.12500}%
\definecolor{mycolor4}{rgb}{0.49400,0.18400,0.55600}%
\definecolor{mycolor5}{rgb}{0.46600,0.67400,0.18800}%
\begin{tikzpicture}

\begin{axis}[%
width=4.5cm,
height=1.5cm,
at={(0.758in,0.481in)},
scale only axis,
xmin=0,
xmax=160,
xlabel style={font=\color{white!15!black}},
ymode=log,
ymin=100,
ymax=1000000000000,
ytick = {100,100000,100000000000},
yticklabels = {$\quad \quad \quad$,$10^5$,$10^{10}$},
yminorticks=true,
ylabel style={font=\color{white!15!black}},
ylabel={Eigenvalue ratio $\rho$},
axis background/.style={fill=white},
legend pos=outer north east,
legend style={legend cell align=left, align=left, draw=white!15!black}
]
\addplot [color=mycolor1]
  table[row sep=crcr]{%
0	5050709256.29674\\
10	1061669406.46793\\
20	97901488160.9708\\
30	4685817637.03288\\
40	4961011183.62816\\
50	535282927.189803\\
60	3417649554.34636\\
70	5293209066.87006\\
80	41841796036.2865\\
90	785889380.48417\\
100	27021121684.7998\\
110	314011338.745289\\
120	19850138074.399\\
130	4078039425.98481\\
140	166862.610880683\\
150	135692.803552457\\
160	142185.65542412\\
170	140434.599768167\\
180	160727.278027547\\
190	150992.640913667\\
200	158594.168298614\\
};
\addlegendentry{$\rho\text{(W}_\text{0}\text{)}$}

\addplot [color=mycolor2]
  table[row sep=crcr]{%
0	2447424403.26092\\
10	586070659.658258\\
20	60902213717.8388\\
30	3387599629.80214\\
40	4242211042.77961\\
50	514068619.214549\\
60	3789542863.23482\\
70	7642569620.36609\\
80	77989266331.8311\\
90	1516583194.03458\\
100	74862948340.5585\\
110	1103536523.47303\\
120	58021318247.5572\\
130	56190128616.5967\\
140	679057174.555919\\
150	1241889284.72768\\
160	1993441403.3011\\
170	2484809284.80475\\
180	949229359.814326\\
190	853208781.189237\\
200	667378369.465936\\
};
\addlegendentry{$\rho (W_0 + \kappa_1 \tilde{B}_1^u)$}

\addplot [color=mycolor3]
  table[row sep=crcr]{%
0	1900.23917725635\\
10	30241765.8761519\\
20	6051204854.71268\\
30	512884199.054442\\
40	768649562.023033\\
50	110856919.522559\\
60	959699944.544223\\
70	2085214011.21173\\
80	22441294004.4177\\
90	504992321.07638\\
100	26245544186.0033\\
110	413579195.80172\\
120	18787183785.6495\\
130	22951739450.949\\
140	282263721.230191\\
150	513364170.780948\\
160	824669808.082833\\
170	1025656551.58294\\
180	384439335.563631\\
190	341240612.580151\\
200	262186182.283602\\
};
\addlegendentry{$\rho (W_0 - \kappa_1 \tilde{B}_1^l)$}

\addplot [color=mycolor4]
  table[row sep=crcr]{%
0	6394420425.77164\\
10	1452207295.69797\\
20	143579218435.288\\
30	7649681091.23976\\
40	9249754524.56176\\
50	1065139169.81288\\
60	7645689892.80661\\
70	14630720085.2965\\
80	141573300903.413\\
90	2462183559.09214\\
100	129507221322.293\\
110	1835368036.87784\\
120	100293914582.114\\
130	83409651878.4575\\
140	1082593189.57021\\
150	1975389539.77455\\
160	3197347149.40176\\
170	4063558408.98223\\
180	1578347674.72689\\
190	1447277850.65209\\
200	1156064806.40759\\
};
\addlegendentry{$\rho (W_0 + \kappa_2 \tilde{B}_2^u)$}

\addplot [color=mycolor5]
  table[row sep=crcr]{%
0	954.424897661313\\
10	10381054.7069984\\
20	2152564037.30703\\
30	161714054.426168\\
40	260732911.606293\\
50	37293622.9120723\\
60	329593881.215844\\
70	697396695.597115\\
80	7798237737.53753\\
90	166349556.665254\\
100	8637229920.5392\\
110	139694924.278187\\
120	7986893467.41655\\
130	8039076669.41141\\
140	94806689.2328173\\
150	194282717.89792\\
160	336648707.310162\\
170	448139256.794565\\
180	182447346.248012\\
190	176895601.452563\\
200	146214124.095046\\
};
\addlegendentry{$\rho (W_0 - \kappa_2 \tilde{B}_2^l)$}
    \addplot [color=red,dashed]
      table[row sep=crcr]{%
      0 100000 \\
      500 100000 \\
      };
\end{axis}
\end{tikzpicture}%
%
%
\definecolor{mycolor1}{rgb}{0.00000,0.44700,0.74100}%
\begin{tikzpicture}

\begin{axis}[%
width=4.3cm,
height=0.75cm,
at={(0.793in,0.481in)},
scale only axis,
xmin=0,
xmax=160,
xlabel style={font=\color{white!15!black}},
ymin=58222,
ymax=58227,
scaled ticks=false,
ytick = {58222,58227},
ylabel style={font=\color{white!15!black}},
ylabel={Cost (\$/h)},
legend pos=outer north east,
axis background/.style={fill=white},
legend style={legend cell align=left,cells={align=left}, align=left, draw=white!15!black}
]
\addplot [color=mycolor1]
  table[row sep=crcr]{%
0	58224.1663200684\\
10	58224.160689846\\
20	58224.1814529938\\
30	58224.1897620516\\
40	58224.1928862424\\
50	58224.1836383782\\
60	58224.200589258\\
70	58224.2062675978\\
80	58224.2130957218\\
90	58224.2139960816\\
100	58224.2249903637\\
110	58224.2248042523\\
120	58224.3760161871\\
130	58224.383558137\\
140	58226.7250383401\\
150	58226.6761907503\\
160	58226.6776179821\\
170	58226.716811834\\
180	58226.7350119675\\
190	58226.7693137988\\
200	58226.7206105521\\
};
\addlegendentry{Gen. \\ cost}

\end{axis}
\end{tikzpicture}%
%
%
\definecolor{mycolor1}{rgb}{0.00000,0.44700,0.74100}%
\begin{tikzpicture}

\begin{axis}[%
width=4.3cm,
height=0.75cm,
at={(0.758in,0.481in)},
scale only axis,
xmin=0,
xmax=160,
xlabel style={font=\color{white!15!black}},
xlabel={Weight for droop penalty $\mu$ (p.u.)},
ymin=-0,
ytick = {0,30,60},
yticklabels = {$\quad \quad \quad$,30,60},
ymax=60,
ylabel style={font=\color{white!15!black}},
ylabel={Cost (\$/h)},
legend pos=outer north east,
axis background/.style={fill=white},
legend style={legend cell align=left, align=left,cells={align=left}, draw=white!15!black}
]
\addplot [color=mycolor1]
  table[row sep=crcr]{%
0	-0\\
10	2.59659553886936\\
20	5.20468860927724\\
30	7.81656550175785\\
40	10.4221214588493\\
50	13.027382354392\\
60	15.6356489193114\\
70	18.2440147606293\\
80	20.8529201135912\\
90	23.4608861152676\\
100	26.0719704736868\\
110	28.6798374553948\\
120	31.2773813636591\\
130	33.8973448355801\\
140	36.500365078162\\
150	39.1160945513472\\
160	41.7273559861374\\
170	44.3397680921024\\
180	46.9244764199694\\
190	49.5215539745508\\
200	52.1467357565289\\
};
\addlegendentry{Penalty$\,\,$\\ term}

\end{axis}
\end{tikzpicture}%
		\end{footnotesize}
	\caption{Eigenvalue ratios $\rho$, generation cost and penalty term as a function of the power loss penalty weight $\mu$ for a IEEE 24 bus test case with two wind farms and a Gaussian uncertainty set.}
	\label{EV_ratio_9Gauss}
\end{figure}
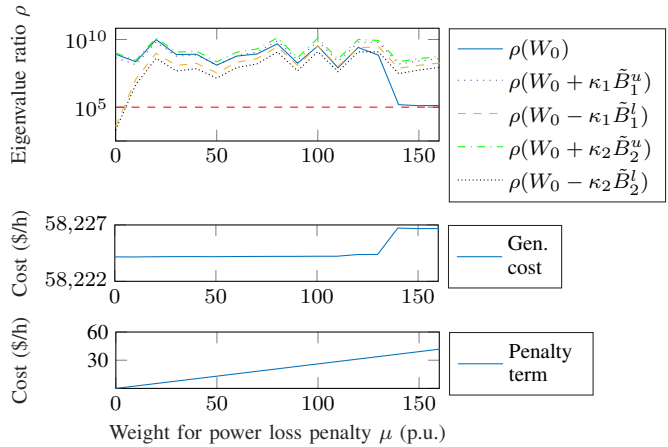

A similar observation can be made if a Gaussian distribution is assumed for the forecast errors. Fig.~\ref{EV_ratio_9Gauss} shows the eigenvalue ratios $\rho$ as a function of the penalty weight $\mu$ for the Gaussian uncertainty set. A certain minimum value for the weight $\mu = 10$ is necessary to obtain solution matrices with \mbox{rank-1} at the operating state $W_0$ and the four end-point of the ellipsoid axes. The generation cost is almost flat with respect to increasing penalty weight and the near-global optimality at $\mu = 10$ for this test case evaluates to an upper bound larger than $99.99\%$. As it is also observed, the necessary magnitude of the penalty weight $\mu$ to obtain \mbox{rank-1} solution matrices depends on the test case and configuration.

\subsection{IEEE 118 Bus Test Case}

In this section, our proposed approaches using the affine policy and PTDFs are compared with two alternative approaches described in the literature \cite{jabr2015robust, schmidli2016stochastic}. We use the IEEE 118 bus test case with realistic forecast data for the wind farms, and Monte Carlo simulations to evaluate the constraint violations.
\subsubsection{Simulation Setup}
\begin{figure}
	\begin{footnotesize}
%
%
\begin{tikzpicture}

\begin{axis}[%
width=5.5cm,
height=2cm,
at={(0.758in,0.481in)},
scale only axis,
xmin=1,
xmax=5,
xlabel style={font=\color{white!15!black}},
ymin=0,
ymax=1,
legend pos=outer north east,
ylabel style={font=\color{white!15!black}},
ylabel={$\text{P}_{\text{W}_1} \text{ (p.u.)}$},
axis background/.style={fill=white},
legend style={legend cell align=left, align=left, draw=white!15!black}
]

\addplot[area legend, draw=white!90!black, fill=white!90!black]
table[row sep=crcr] {%
x	y\\
1   0.81441\\
2   0.88887\\
3   0.87628\\
4   0.99607\\
5   0.97444\\
5   0.27439\\
4   0.32884\\
3   0.40938\\
2   0.34953\\
1   0.33971\\
}--cycle;
\addlegendentry{Bounds}

\addplot [color=blue, mark=x, mark options={solid, blue}]
  table[row sep=crcr]{%
1   0.55057\\
2   0.61216\\
3   0.68057\\
4   0.7448\\
5   0.76123\\
};
\addlegendentry{Forecast}

\end{axis}
\end{tikzpicture}%
%
%
\begin{tikzpicture}

\begin{axis}[%
width=5.5cm,
height=2cm,
at={(0.758in,0.481in)},
scale only axis,
xmin=1,
xmax=5,
xlabel style={font=\color{white!15!black}},
xlabel={Time (h)},
ymin=0,
ymax=1,
legend pos=outer north east,
ylabel style={font=\color{white!15!black}},
ylabel={$\text{P}_{{\text{W}_2}} \text{ (p.u.)}$},
axis background/.style={fill=white},
legend style={legend cell align=left, align=left, draw=white!15!black}
]

\addplot[area legend, draw=white!90!black, fill=white!90!black]
table[row sep=crcr] {%
x	y\\
1 0.67064\\
2 0.71668\\
3 0.84221\\
4 0.94187\\
5 0.87072\\
5 0.26692\\
4 0.45222\\
3 0.40605\\
2 0.34654\\
1 0.2361\\
}--cycle;
\addlegendentry{Bounds}

\addplot [color=blue, mark=x, mark options={solid, blue}]
  table[row sep=crcr]{%
1 0.46179\\
2 0.54078\\
3 0.61234\\
4 0.65672\\
5 0.66273\\
};
\addlegendentry{Forecast}

\end{axis}
\end{tikzpicture}%
	\end{footnotesize}
	\caption{Forecast data from hour 1 to hour 5. The bounds correspond to the minimum and maximum values from the $N_s$ sampled scenarios for the rectangular uncertainty set.}
	\label{ForecastData}
\end{figure}
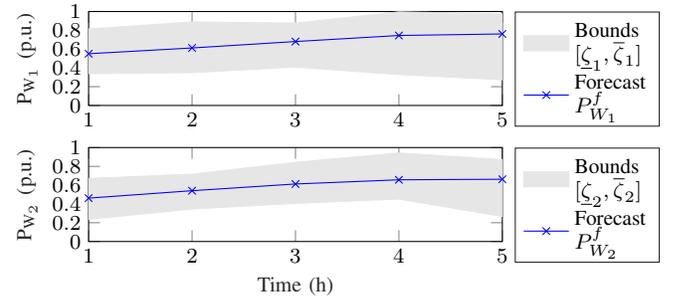
\begin{figure}
	\begin{footnotesize}
		\input{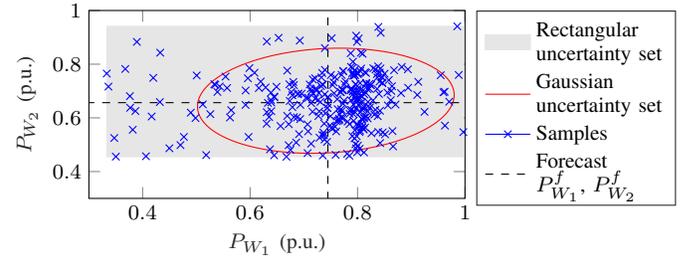}
	\end{footnotesize}
	\caption{Comparison of rectangular and Gaussian uncertainty set obtained from the $N_s$ scenarios sampled for hour 4.}
	\label{UncertaintySets}
\end{figure}
We use the IEEE 118 bus specifications from \cite{IEEE118bus} with the following modifications: The bus voltage limits are set to 0.94 p.u. and 1.06 p.u. As the upper branch flow limits are specified in MW, the active line flow limit is considered for branch flows. The line flow limits are decreased by 30\% and the load is increased by 30\% to obtain a more constrained system. Two wind farms with a rated power of 300 MW and 600 MW are placed at buses 5 and 64. The rated wind power corresponds to 24.1\% of total load demand. Realistic day-ahead wind forecast scenarios from \cite{pinson2013wind} and \cite{Bukhsh2016} are used for both wind farms. To create the scenarios, the methodology described in \cite{pinson2013wind} is used. The forecasts are based on wind power measurements in the Western Denmark area from 15 different control zones collected by the Danish transmission system operator Energinet. We select control zone 1 to correspond to the wind farm at bus 5 and zone 7 to the wind farm at bus 64. We allow a constraint violation of $\epsilon = 5\%$ for all considered approaches. In order to construct the rectangular uncertainty set, the confidence parameter $\beta = 10^{-3}$ is selected. Then, a minimum of 314 scenarios are required according to \eqref{NumberOfSamples}. The forecast is computed as mean value of the scenarios. For the Gaussian uncertainty set, we compute the co-variance matrix based on these 314 scenarios. Fig.~\ref{ForecastData} shows the forecast data from hour 1 to hour 5 with the upper and lower bounds specified by the maximum and minimum scenario values, respectively. In Fig.~\ref{UncertaintySets} the rectangular and Gaussian uncertainty set for hour 4 are shown.

In the following, the parameters for the corrective control policies are specified. A participation factor of $0.25$ is selected for the generators at buses $i = \{12, 26,54, 61\}$, i.e. $d_{G_i} = 0.25$. Wind farms have a reactive power capability of 0.95 inductive to 0.95 capacitive according to recent Grid Codes \cite{tsili2009review}. The approaches using PTDFs assign a fixed power factor $\cos\phi$ to each wind farm. The affine policy includes a generator voltage and wind farm reactive power corrective control, assigning an updated set-point to generators and wind farms based on the actual realization of the forecast errors. 
To facilitate comparability, we use the same scenarios for all approaches to compute the respective uncertainty sets. We evaluate the constraint violations using Monte Carlo simulations with 10'000 scenarios and MATPOWER AC power flows \cite{zimmerman2011matpower}. We enable the enforcement of generator reactive power limits in the power flow, i.e. $PV$ buses are converted to $PQ$ buses once the limits are reached, as otherwise high nonphysical overloading of the limits can occur \cite{Efthymiadis1996}. Furthermore, we distribute the loss mismatch from the active generator set-points among the generators according to their participation factors and rerun the power flow to mimic the response of automatic generation control (AGC).\\

\subsubsection{Numerical Comparison to Alternative Approaches}
In the following, the main modeling assumptions of the respective approaches and the type of chance constraints they include are outlined. All approaches considering chance constraints include corrective control of the active generator set-points.
\begin{itemize}
	\item Chance constrained DC-OPF \cite{jabr2015robust} (DC-OPF): A robust formulation based on DC-OPF includes chance constraints on active generator power and active branch flow. Interval bounds on the forecast errors are assumed. Hence, we use the scenarios to compute the interval bounds. A power factor of 1 is assumed for wind farms. 
	\item Iterative chance constrained AC-OPF \cite{schmidli2016stochastic} (Iterative): At each iteration the Jacobian is computed and the uncertainty margins resulting from the chance constraints are updated until convergence is reached. The forecast errors are assumed to follow a Gaussian distribution. The covariance matrix constructed from the $N_s$ scenarios is used. Chance constraints on active and reactive generator limits, voltage magnitudes and apparent line flows are included in the formulation. A power factor of 1 is assumed for wind farms, as no reactive power corrective control is included in \cite{schmidli2016stochastic}.
\end{itemize}
These two approaches are compared to the following approaches based on the formulations presented in this work:
\begin{itemize}
    \item AC-OPF with convex relaxations but without chance constraints (AC-OPF) \cite{lavaei2012zero}
    \item Chance constrained AC-OPF with convex relaxations, using an affine policy for a Gaussian uncertainty set (AP (Gauss)) including corrective control for wind farms, generator voltages and generator active power.
	\item Chance constrained AC-OPF with convex relaxations, using an affine policy for a rectangular uncertainty set (AP (Rect)) including corrective control for wind farms and generator voltages and generator active power.
	\item Chance constrained AC-OPF with convex relaxations, using PTDFs (PTDF (Gauss))  for a Gaussian uncertainty set.
	\item Chance constrained AC-OPF with convex relaxations, using PTDFs (PTDF (Rect)) for a rectangular uncertainty set.
\end{itemize}
\begin{table}[]
    \centering
    \caption{Cost of uncertainty: Generation cost in relation to AC-OPF with convex relaxations without consideration of uncertainty}
    \begin{tabular}{l r r r r r}
    \toprule
    \multicolumn{1}{l}{Time step (h)} & \multicolumn{1}{c}{1} & \multicolumn{1}{c}{2} & \multicolumn{1}{c}{3} & \multicolumn{1}{c}{4} & \multicolumn{1}{c}{5} \\
    \midrule
   AP (Rect) (\%) &0.774  &  0.740  &  0.755 &   0.921 &  1.562 \\
   \midrule
   PTDF (Rect) (\%)  & 0.785  &  0.748  &  0.767 &   0.931  &  1.588 \\
   \midrule
   DC-OPF \cite{jabr2015robust} (\%)  & -2.782  & -2.826  & -2.824 &  -2.673 &  -2.165 \\
   \midrule
   AP (Gauss) (\%)  & 0.515  &  0.461  &  0.467 &   0.489 &  0.512 \\
   \midrule
   PTDF (Gauss) (\%)  &  0.523  &  0.468  &  0.477 &   0.497 &  0.520 \\
   \midrule
   Iterative \cite{schmidli2016stochastic} (\%)   & 0.519  &  0.465  &  0.473 &   0.494 &  0.516 \\
   \bottomrule
    \end{tabular}
    \label{GenCost188}
\end{table}
In Table~\ref{GenCost188} the cost of uncertainty for the different approaches and considered time steps are shown. The cost of uncertainty represents the additional cost incurred by considering the stochastic variables, and is defined as the difference between the solution of the chance constrained and a baseline. In this paper, the AC-OPF with convex relaxations but without considering uncertainty is assumed as the baseline cost. From Table~\ref{GenCost188}, we make the following observations. First, the DC-OPF (with chance constraints) leads to a cost reduction, as no losses are considered compared to the AC-OPF. Second, the approaches stemming from robust optimization lead to a cost increase of approximately 0.8\% for time step 1 compared to an increase of approximately 0.5\% for the same time step for the approaches assuming a Gaussian distribution. This shows that the Gaussian uncertainty set is less conservative as indicated in Fig.~\ref{UncertaintySets}. For the rectangular uncertainty set, the affine policy reduces the cost compared to the approach using PTDFs. Comparing the approaches for the Gaussian uncertainty set, again the affine policy results to the lowest cost of uncertainty compared to the approach using PTDFs and the iterative chance constrained AC-OPF. The reason for that is that the affine policy includes corrective control for voltages and both active and reactive power.

\begin{table}
		\caption{Violation probability of the chance constraints on bus voltage, active branch flow and active generator power for the forecast data. Monte Carlo simulations using 10'000 scenarios with MATPOWER AC power flows are conducted. Insecure instances are marked in bold.}
	\label{MonteCarloAnalysis}
	\centering
    \begin{tabular}{l c c c c c}
    \toprule
    \toprule
    Time step (h) & 1 & 2 & 3 & 4 & 5 \\
    \midrule
    \midrule
    \multicolumn{6}{c}{Bus voltage}\\
    \midrule
    AC-OPF (\%) & 0.0   & 0.1  &   0.2 &    1.0   &  4.3\\
    \midrule
       DC-OPF \cite{jabr2015robust} (\%)  & \textbf{100.0}  & \textbf{100.0} &  \textbf{100.0}  & \textbf{100.0} &  \textbf{100.0} \\
   \midrule
   PTDF (Rect) (\%)  & \textbf{19.5}  & \textbf{20.9}  &  \textbf{14.6}  &  \textbf{13.0} &   \textbf{12.9}\\
   \midrule
   AP (Rect) (\%) &0.0   &  0.0    &  0.0    & 0.0  &   0.0 \\
   \midrule
   PTDF (Gauss) (\%)  &  \textbf{20.0} & \textbf{21.2}   & \textbf{15.0}   & \textbf{13.5}&    \textbf{13.0}\\
   \midrule
      AP (Gauss) (\%) & 0.7    &1.9  &   4.3    & \textbf{7.2}   &  \textbf{7.6} \\
   \midrule
   Iterative \cite{schmidli2016stochastic} (\%)   & 0.0   & 0.0     &0.0  &   0.0   &  0.0 \\
   \midrule
   \midrule 
       \multicolumn{6}{c}{Active power line limit}\\
    \midrule
   AC-OPF (\%) &\textbf{17.7}   &\textbf{18.8}   & \textbf{14.9} &   \textbf{32.5} &   \textbf{46.5}\\
      \midrule
      DC-OPF \cite{jabr2015robust} (\%)  &0.0  &  0.0   &  0.0   &  2.8 &    0.0 \\
   \midrule
   PTDF (Rect) (\%)  & 0.0  &  0.0  &   0.0 &    0.0   &  0.0 \\
      \midrule
   AP (Rect) (\%) & 0.0  &  0.0   &  0.0 &    0.0   &  0.0 \\
   \midrule
   PTDF (Gauss) (\%) &  4.6  &  \textbf{11.1}  &  \textbf{13.1} &   \textbf{9.3}   &  \textbf{7.8} \\
      \midrule
      AP (Gauss) (\%) & 4.6   & 3.7  &   0.9 &    2.6    & \textbf{5.8}\\
   \midrule
   Iterative \cite{schmidli2016stochastic} (\%) &  1.6   & 2.0   &  3.6  &   4.2    & \textbf{5.3} \\
   \midrule 
   \midrule 
          \multicolumn{6}{c}{Active generator limit}\\
    \midrule
   AC-OPF (\%) &\textbf{46.4}   & \textbf{48.8} &   \textbf{45.9}   & \textbf{45.5}  &  \textbf{40.9} \\
    \midrule
      DC-OPF \cite{jabr2015robust} (\%)  & \textbf{34.3}  &  \textbf{38.6}   & \textbf{30.1} &   \textbf{14.5} &   2.0 \\
      \midrule
   PTDF (Rect)  &0.0   &  0.0     &0.0    & 0.0     &0.0\\
   \midrule
   AP (Rect) (\%) & 0.0    & 0.0 &    0.0  &   0.0 &    0.0 \\
      \midrule
   PTDF (Gauss) (\%) & 2.6   &  3.6   &  2.8  &  3.1    & \textbf{5.5} \\
     \midrule
   AP (Gauss) (\%)  &0.0  &   0.2 &    0.4  &   2.3 &    0.7 \\
   \midrule
   Iterative \cite{schmidli2016stochastic} (\%)   & 2.9   & 4.1  &   3.0     &3.3    & \textbf{5.7} \\
   \bottomrule
   \bottomrule
    \end{tabular}
\end{table}
In Table~\ref{MonteCarloAnalysis} the violation probability of the chance constraints on active power, voltages, and active branch flows are shown. Monte Carlo simulations using 10'000 scenarios with MATPOWER AC power flows are conducted. A minimum violation limit of $10^{-3}$ p.u. for active generator limits and 0.1\% for voltage and line flow limits is considered to exclude numerical errors. In all considered time steps, the AC-OPF without consideration of uncertainty leads to insecure instances and violates constraints on line and generator limits on active power.

First, investigating the robust approaches using the rectangular uncertainty set the following observations can be made: The robust DC-OPF formulation in \cite{jabr2015robust} leads to insecure instances for all time steps and violates both voltage and generator active power constraints. The AC-OPF approach using PTDFs for the chance constraints reduces the voltage violations but does not comply with the 5\% confidence interval. The AC-OPF using the affine policy complies with the chance constraints for all time steps while slightly decreasing the generation cost compared to the approach using PTDFs. As the scenario based method is conservative, there are nearly zero violations occurring for the considered 10'000 samples for the approach using the affine policy.

Second, we compare the different approaches which assume a Gaussian distribution of the forecast errors. The affine policy improves upon the approach using PTDFs and results to a secure operation for time steps 1 to 3. For time steps 4 and 5 we observe a slight violation of the active power line and bus voltage limit. This is due to the fact that we do not sample out of a Gaussian distribution but out of a set of realistic forecast scenarios, that apparently are not Gaussian distributed. The iterative approach results to a secure operation for time steps 1 to 4 and slightly violates the active generator and branch flow limit in time step 5.

\begin{table}
		\caption{Comparison of violation probability of the chance constraints on bus voltage, active branch flow and active generator power for affine policy and iterative AC-OPF with 10'000 samples from a Gaussian distribution.}
	\label{SampledGauss}
	\centering
    \begin{tabular}{l c c c c c}
    \toprule
    \toprule
    Time step (h) & 1 & 2 & 3 & 4 & 5 \\
    \midrule
    \midrule
    \multicolumn{6}{c}{Bus voltage}\\
   \midrule
   AP (Gauss) (\%) & 0.1 &  0.4 & 0.5 &  0.8 & 0.7  \\
   \midrule
   Iterative \cite{schmidli2016stochastic} (\%)   & 0.0  & 0.0   & 0.0  &  0.0  &  0.0 \\
   \midrule
   \midrule 
       \multicolumn{6}{c}{Active power line limit}\\
    \midrule
      AP (Gauss) (\%) &  2.2  & 1.7 &  2.1  &  2.1   & 2.1\\
   \midrule
   Iterative \cite{schmidli2016stochastic} (\%) &  1.5 & 1.7  &  1.7 &  2.1 &  2.1\\
   \midrule 
   \midrule 
          \multicolumn{6}{c}{Active generator limit}\\
    \midrule
   AP (Gauss) (\%)  &  2.4 &  2.7  &   2.7  & 2.6  &   2.5 \\
   \midrule
   Iterative \cite{schmidli2016stochastic} (\%)   & 2.7 & 2.6  &  2.3  & 2.4 & 2.1 \\
   \bottomrule
   \bottomrule
    \end{tabular}
\end{table}
In order to verify if these violations occur due to the mismatch between actual distribution and the assumed Gaussian we repeat the 10'000 scenario evaluations for both affine policy and the iterative chance constrained AC-OPF from \cite{schmidli2016stochastic}. We sample from the Gaussian distribution assumed for the uncertainty set. The results are shown in Table~\ref{SampledGauss}. For all 5 time steps, both approaches comply with the $5\%$ violation probability. Hence, the occurring violations in Table~\ref{MonteCarloAnalysis} stem from the mismatch between Gaussian distribution and actual probability distribution. As shown in Table~\ref{GenCost188}, the affine policy results in a slightly lower generation cost than the iterative AC-OPF, as it includes corrective control policies. This leads us to the following conclusions. First, that if the forecast errors do follow a normal distribution both approaches demonstrate good performance and do not exceed the violation limit. If the data are not normally distributed, as is the case for the results shown in Table~\ref{MonteCarloAnalysis}, none of the two methods can guarantee that the violation probability will be below $\epsilon$. The differences in performance in that case are, as it would be expected, data- and system-specific. However, independent from the fact if the underlying probability distribution is Gaussian or not, one difference that remains is that the approach proposed in this paper is more rigorous, since it provides guarantees regarding the global optimality of the obtained solution and allows to include corrective control policies related to reactive power and voltage.

\begin{table}
	\begin{center}
		\caption{Power loss penalty weight and near-global optimality guarantees for IEEE 118 bus test case for all considered time steps}
		\label{GlobOpt}
		\begin{tabular}{l c c}
			\toprule
			& Penalty weight $\mu$ & Near-global optimality \\
			& (p.u.) & guarantee $\delta_{\text{opt}}$ (\%) \\
			\midrule
			 AP (Rect) & 100  & $\geq 99.99 $ \\
			 \midrule
			 AP (Gauss) & 100 & $\geq 99.99 $ \\
			\bottomrule
		\end{tabular}
	\end{center}	
\end{table}
Table~\ref{GlobOpt} lists the penalty weights and obtained near-global optimality guarantees for the 5 time steps. Note that it is sufficient to define for both uncertainty sets a penalty weight of $\mu = 100$ p.u. to obtain zero relaxation gap, i.e. \mbox{rank-1} solution matrices and a near-global optimality guarantee of larger than 99.99\%. This means that the maximum deviation from the global optimum is smaller than 0.01\% of the objective value.

\begin{table}
	\begin{center}
		\caption{Solving time for IEEE 118 bus test case}
		\label{CompTime}
		\begin{tabular}{c c c}
			\toprule
			AP (Rect) & AP (Gauss) & PTDF (Rect) \\
			\midrule
			30 sec  & 10 min  &  15 sec  \\
			\midrule
			PTDF (Gauss) & DC-OPF \cite{jabr2015robust} & Iterative \cite{schmidli2016stochastic} \\
			\midrule
			15   sec  & $ \leq$ 1  sec  &       4 sec  \\
			\bottomrule
		\end{tabular}
	\end{center}
\end{table}
Table~\ref{CompTime} lists the computational time of the different approaches. The optimization problems are solved on a desktop computer with an Intel Xeon CPU E5-1650 v3 @ 3.5 GHz and 32 GB RAM. For all optimization problems except the iterative approach, MOSEK V8 \cite{mosek} is used. The iterative approach utilizes the MATPOWER AC-OPF. The DC-OPF formulation is the fastest, as the optimization problem is a linear program. The computational time increases with increasing constraint complexity. The SOC constraints in the formulation for the Gaussian uncertainty set are computationally the most challenging. We observe that the iterative approach, despite the need for computing a number of iterations, converges faster than all approaches that utilize convex relaxations and an SDP solver. Current trends expect the need of more rigorous optimal power flow approaches in the future, that e.g. can guarantee a global minimum. In that case the need for further research to improve both the optimization solvers and the convex formulations of the AC-OPF problem is apparent. Possible directions to increase the computational speed of the proposed approaches are the chordal decomposition technique, outlined in \cite{molzahn2013implementation}, and distributed optimization techniques, e.g. the alternating direction method of multipliers (ADMM) for sparse semidefinite problems in \cite{Madani2015ADMM}. The chordal decomposition technique can be applied to reduce the computational burden of the semidefinite constraints on $W_0$ and $B_i$ \eqref{SDP}. As shown in \cite{jabr2012exploiting} a speed-up by several orders of magnitude can then be expected for large systems.
\section{Conclusions}\label{VI}
In this work, a convex formulation for a chance constrained AC-OPF is presented which is able to provide near-global optimality guarantees. The OPF formulation considers chance constraints for all relevant variables, and has an explicit representation of corrective control policies. Two tractable formulations are proposed: First, a scenario-based method is applied in combination with robust optimization. Second, assuming a Gaussian distribution of forecast errors, we provide an analytical reformulation of the chance constraints. Detailed case studies on the IEEE 24 and 118 bus test systems are presented. For the latter, we used realistic forecast data and Monte Carlo simulations to evaluate constraint violations. Compared to a chance constrained DC-OPF formulation, we find that the formulations proposed in this paper are more accurate and significantly decrease constraint violations. Compared with iterative non-convex AC-OPF formulations, both our piece-wise affine control policy and the iterative AC-OPF do not exceed the constraint violation limit for the Gaussian uncertainty set. Most importantly, our proposed approach obtains tight near-global optimality guarantees which ensure that the distance to the global optimum is smaller than 0.01\% of the objective value. In our future work, besides investigating chordal decomposition techniques, we include security constraints in the proposed formulation by defining a matrix $W^s(\zeta)$ for each outage $s$ of a generation unit or transmission line.
\section*{Acknowledgment}
The authors would like to thank Pierre Pinson for sharing the forecast data, Line Roald for providing an updated version of the code from \cite{schmidli2016stochastic}, and Martin S. Andersen and Daniel K. Molzahn for fruitful discussions.
\bibliographystyle{IEEEtran}

\bibliography{CCACOPF}

\begin{thebibliography}{10}
\providecommand{\url}[1]{#1}
\csname url@samestyle\endcsname
\providecommand{\newblock}{\relax}
\providecommand{\bibinfo}[2]{#2}
\providecommand{\BIBentrySTDinterwordspacing}{\spaceskip=0pt\relax}
\providecommand{\BIBentryALTinterwordstretchfactor}{4}
\providecommand{\BIBentryALTinterwordspacing}{\spaceskip=\fontdimen2\font plus
\BIBentryALTinterwordstretchfactor\fontdimen3\font minus
  \fontdimen4\font\relax}
\providecommand{\BIBforeignlanguage}[2]{{%
\expandafter\ifx\csname l@#1\endcsname\relax
\typeout{** WARNING: IEEEtran.bst: No hyphenation pattern has been}%
\typeout{** loaded for the language `#1'. Using the pattern for}%
\typeout{** the default language instead.}%
\else
\language=\csname l@#1\endcsname
\fi
#2}}
\providecommand{\BIBdecl}{\relax}
\BIBdecl

\bibitem{nemirovski2006convex}
A.~Nemirovski and A.~Shapiro, ``Convex approximations of chance constrained
  programs,'' \emph{SIAM Journal on Optimization}, vol.~17, no.~4, pp.
  969--996, 2006.

\bibitem{bienstock2014chance}
D.~Bienstock, M.~Chertkov, and S.~Harnett, ``Chance-constrained optimal power
  flow: Risk-aware network control under uncertainty,'' \emph{SIAM Review},
  vol.~56, no.~3, pp. 461--495, 2014.

\bibitem{roald2013analytical}
L.~Roald, F.~Oldewurtel, T.~Krause, and G.~Andersson, ``Analytical
  reformulation of security constrained optimal power flow with probabilistic
  constraints,'' in \emph{IEEE PowerTech}, Grenoble, France, 2012.

\bibitem{lubin2015robust}
M.~Lubin, Y.~Dvorkin, and S.~Backhaus, ``A robust approach to chance
  constrained optimal power flow with renewable generation,'' \emph{IEEE
  Transactions on Power Systems}, vol.~31, no.~5, pp. 3840 -- 3849, 2016.

\bibitem{jabr2015robust}
R.~A. Jabr, S.~Karaki, and J.~A. Korbane, ``Robust multi-period {OPF} with
  storage and renewables,'' \emph{IEEE Transactions on Power Systems}, vol.~30,
  no.~5, pp. 2790--2799, 2015.

\bibitem{roald2017corrective}
L.~Roald, S.~Misra, T.~Krause, and G.~Andersson, ``Corrective control to handle
  forecast uncertainty: A chance constrained optimal power flow,'' \emph{IEEE
  Transactions on Power Systems}, vol.~32, no.~2, pp. 1626--1637, 2017.

\bibitem{guggilam2015scalable}
S.~S. Guggilam, E.~Dall'Anese, Y.~C. Chen, S.~V. Dhople, and G.~B. Giannakis,
  ``Scalable optimization methods for distribution networks with high {PV}
  integration,'' \emph{IEEE Transactions on Smart Grid}, vol.~7, no.~4, pp.
  2061 -- 2070, 2016.

\bibitem{baker2016distribution}
K.~Baker, E.~Dall'Anese, and T.~Summers, ``Distribution-agnostic stochastic
  optimal power flow for distribution grids,'' in \emph{North American Power
  Symposium (NAPS)}, Denver, US, 2016.

\bibitem{summers2014}
T.~Summers, J.~Warrington, M.~Morari, and J.~Lygeros, ``Stochastic optimal
  power flow based on convex approximations of chance constraints,'' in
  \emph{2014 Power Systems Computation Conference}, 2014.

\bibitem{anese2017chance}
E.~D. Anese, K.~Baker, and T.~Summers, ``Chance-constrained {AC} optimal power
  flow for distribution systems with renewables,'' \emph{IEEE Transactions on
  Power Systems}, vol.~PP, no.~99, pp. 1--1, 2017.

\bibitem{zhang2011chance}
H.~Zhang and P.~Li, ``Chance constrained programming for optimal power flow
  under uncertainty,'' \emph{IEEE Transactions on Power Systems}, vol.~26,
  no.~4, pp. 2417--2424, 2011.

\bibitem{schmidli2016stochastic}
J.~Schmidli, L.~Roald, S.~Chatzivasileiadis, and G.~Andersson, ``Stochastic
  {AC} optimal power flow with approximate chance-constraints,'' in \emph{IEEE
  Power and Energy Society General Meeting}, Boston, US, 2016.

\bibitem{Dvijotham2016}
K.~Dvijotham and D.~K. Molzahn, ``Error bounds on the {DC} power flow
  approximation: A convex relaxation approach,'' in \emph{IEEE 55th Conference
  on Decision and Control (CDC)}, 2016, pp. 2411--2418.

\bibitem{Dhople2015}
S.~V. Dhople, S.~S. Guggilam, and Y.~C. Chen, ``Linear approximations to {AC}
  power flow in rectangular coordinates,'' in \emph{53rd Annual Allerton
  Conference on Communication, Control, and Computing}, 2015, pp. 211--217.

\bibitem{lavaei2012zero}
J.~Lavaei and S.~H. Low, ``Zero duality gap in optimal power flow problem,''
  \emph{IEEE Transactions on Power Systems}, vol.~27, no.~1, pp. 92--107, 2012.

\bibitem{jabr2007conic}
R.~A. Jabr, ``A conic quadratic format for the load flow equations of meshed
  networks,'' \emph{IEEE Transactions on Power Systems}, vol.~22, no.~4, pp.
  2285--2286, 2007.

\bibitem{bai2008semidefinite}
X.~Bai, H.~Wei, K.~Fujisawa, and Y.~Wang, ``Semidefinite programming for
  optimal power flow problems,'' \emph{International Journal of Electrical
  Power \& Energy Systems}, vol.~30, no.~6, pp. 383--392, 2008.

\bibitem{Cane2012ACOPFOneill}
M.~B. Cane, R.~P. O'Neill, and A.~Castillo, ``History of optimal power flow and
  formulations,'' Federal Energy Regulatory Commission, 2012.

\bibitem{Molzahn2011}
B.~C. Lesieutre, D.~K. Molzahn, A.~R. Borden, and C.~L. DeMarco, ``Examining
  the limits of the application of semidefinite programming to power flow
  problems,'' in \emph{49th Annual Allerton Conference on Communication,
  Control, and Computing}, 2011, pp. 1492--1499.

\bibitem{madani2015convex}
R.~Madani, S.~Sojoudi, and J.~Lavaei, ``Convex relaxation for optimal power
  flow problem: Mesh networks,'' \emph{IEEE Transactions on Power Systems},
  vol.~30, no.~1, pp. 199--211, 2015.

\bibitem{vrakopoulou2013probabilistic}
M.~Vrakopoulou, M.~Katsampani, K.~Margellos, J.~Lygeros, and G.~Andersson,
  ``Probabilistic security-constrained {AC} optimal power flow,'' in \emph{IEEE
  PowerTech}, Grenoble, France, 2012.

\bibitem{margellos2014road}
K.~Margellos, P.~Goulart, and J.~Lygeros, ``On the road between robust
  optimization and the scenario approach for chance constrained optimization
  problems,'' \emph{IEEE Transactions on Automatic Control}, vol.~59, no.~8,
  pp. 2258--2263, 2014.

\bibitem{molzahn2013implementation}
D.~K. Molzahn, J.~T. Holzer, B.~C. Lesieutre, and C.~L. DeMarco,
  ``Implementation of a large-scale optimal power flow solver based on
  semidefinite programming,'' \emph{IEEE Transactions on Power Systems},
  vol.~28, no.~4, pp. 3987--3998, 2013.

\bibitem{ben2008selected}
A.~Ben-Tal and A.~Nemirovski, ``Selected topics in robust convex
  optimization,'' \emph{Mathematical Programming}, vol. 112, no.~1, pp.
  125--158, 2008.

\bibitem{Ibraheem2005}
Ibraheem, P.~Kumar, and D.~P. Kothari, ``Recent philosophies of automatic
  generation control strategies in power systems,'' \emph{IEEE Transactions on
  Power Systems}, vol.~20, no.~1, pp. 346--357, 2005.

\bibitem{vu1996improved}
H.~Vu, P.~Pruvot, C.~Launay, and Y.~Harmand, ``An improved voltage control on
  large-scale power system,'' \emph{IEEE Transactions on Power Systems},
  vol.~11, no.~3, pp. 1295--1303, 1996.

\bibitem{tsili2009review}
M.~Tsili and S.~Papathanassiou, ``A review of grid code technical requirements
  for wind farms,'' \emph{IET Renewable Power Generation}, vol.~3, no.~3, pp.
  308--332, 2009.

\bibitem{nemirovski2012safe}
A.~Nemirovski, ``On safe tractable approximations of chance constraints,''
  \emph{European Journal of Operational Research}, vol. 219, no.~3, pp.
  707--718, 2012.

\bibitem{chatzivasileiadis2011flexible}
S.~Chatzivasileiadis, T.~Krause, and G.~Andersson, ``Flexible {AC} transmission
  systems ({FACTS}) and power system security - {A} valuation framework,'' in
  \emph{IEEE Power and Energy Society General Meeting}, Detroit Michigan, US,
  2011.

\bibitem{lubin2015computing}
M.~Lubin and I.~Dunning, ``Computing in operations research using {J}ulia,''
  \emph{INFORMS Journal on Computing}, vol.~27, no.~2, pp. 238--248, 2015.

\bibitem{mosek}
{MOSEK ApS}, \emph{MOSEK 8.0.0.37}, 2016.

\bibitem{zimmerman2011matpower}
R.~D. Zimmerman, C.~E. Murillo-S{\'a}nchez, and R.~J. Thomas, ``{MATPOWER}:
  Steady-state operations, planning, and analysis tools for power systems
  research and education,'' \emph{IEEE Transactions on Power Systems}, vol.~26,
  no.~1, pp. 12--19, 2011.

\bibitem{IEEE118bus}
\BIBentryALTinterwordspacing
``{IEEE} 118-bus, 54-unit, 24-hour system,'' {Electrical and Computer
  Engineering Department, Illinois Institute of Technology}, Tech. Rep.
  [Online]. Available: \url{http://motor.ece.iit.edu/data/JEAS_IEEE118.doc}
\BIBentrySTDinterwordspacing

\bibitem{pinson2013wind}
P.~Pinson, ``Wind energy: Forecasting challenges for its operational
  management,'' \emph{Statistical Science}, pp. 564--585, 2013.

\bibitem{Bukhsh2016}
W.~A. Bukhsh, C.~Zhang, and P.~Pinson, ``An integrated multiperiod {OPF} model
  with demand response and renewable generation uncertainty,'' \emph{IEEE
  Transactions on Smart Grid}, vol.~7, no.~3, pp. 1495--1503, 2016.

\bibitem{Efthymiadis1996}
A.~E. Efthymiadis and Y.~H. Guo, ``Generator reactive power limits and voltage
  stability,'' in \emph{Fourth International Conference on Power System Control
  and Management (Conf. Publ. No. 421)}, 1996, pp. 196--199.

\bibitem{Madani2015ADMM}
R.~Madani, A.~Kalbat, and J.~Lavaei, ``{ADMM} for sparse semidefinite
  programming with applications to optimal power flow problem,'' in \emph{54th
  IEEE Conference on Decision and Control (CDC)}, 2015, pp. 5932--5939.

\bibitem{jabr2012exploiting}
R.~A. Jabr, ``Exploiting sparsity in {SDP} relaxations of the {OPF} problem,''
  \emph{IEEE Transactions on Power Systems}, vol.~27, no.~2, pp. 1138--1139,
  2012.

\end{thebibliography}

\ifCLASSOPTIONcaptionsoff
\newpage
\fi

\begin{IEEEbiography}[{\includegraphics[width=1in,height=1.25in,clip,keepaspectratio]{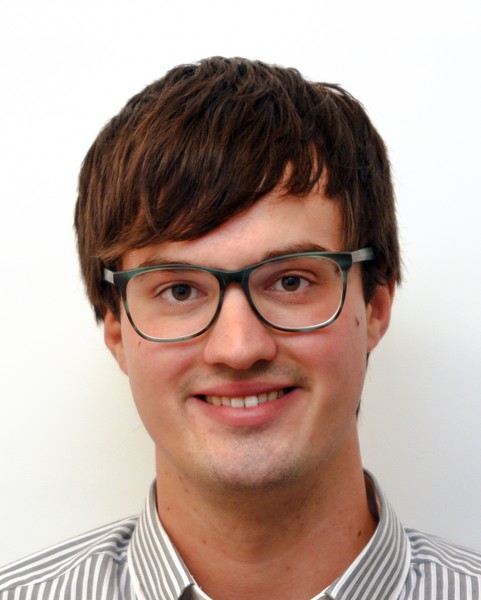}}]{Andreas Venzke} (S'16) received the M.Sc. degree in Energy Science and Technology from ETH Zurich, Zurich, Switzerland in 2017. He is currently working towards the Ph.D. degree at the Department of Electrical Engineering, Technical University of Denmark (DTU), Kongens Lyngby, Denmark. His research interests include power system operation under uncertainty and convex relaxations of optimal power flow.
\end{IEEEbiography}
\begin{IEEEbiography}[{\includegraphics[width=1in,height=1.25in,clip,keepaspectratio]{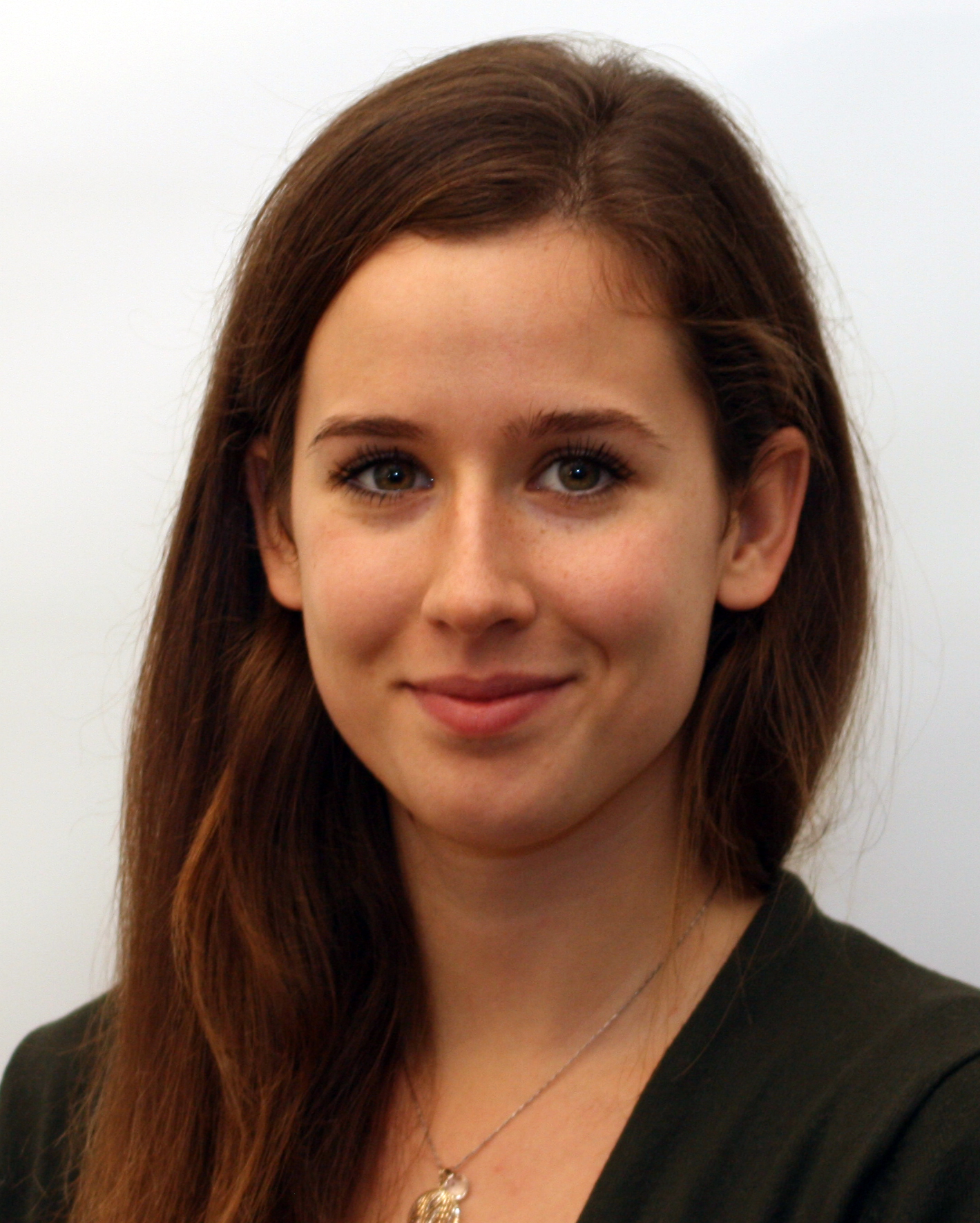}}]{Lejla Halilbasic} (S'15) received the M.Sc. degree in Electrical Engineering from the Technical University of Graz, Austria in 2015. She is currently working towards the Ph.D. degree at the Department of Electrical Engineering, Technical University of Denmark (DTU), Denmark. Her research interests include optimization of power system operation and its applications to electricity markets. 
\end{IEEEbiography}
\begin{IEEEbiography}[{\includegraphics[width=1in,height=1.25in,clip,keepaspectratio]{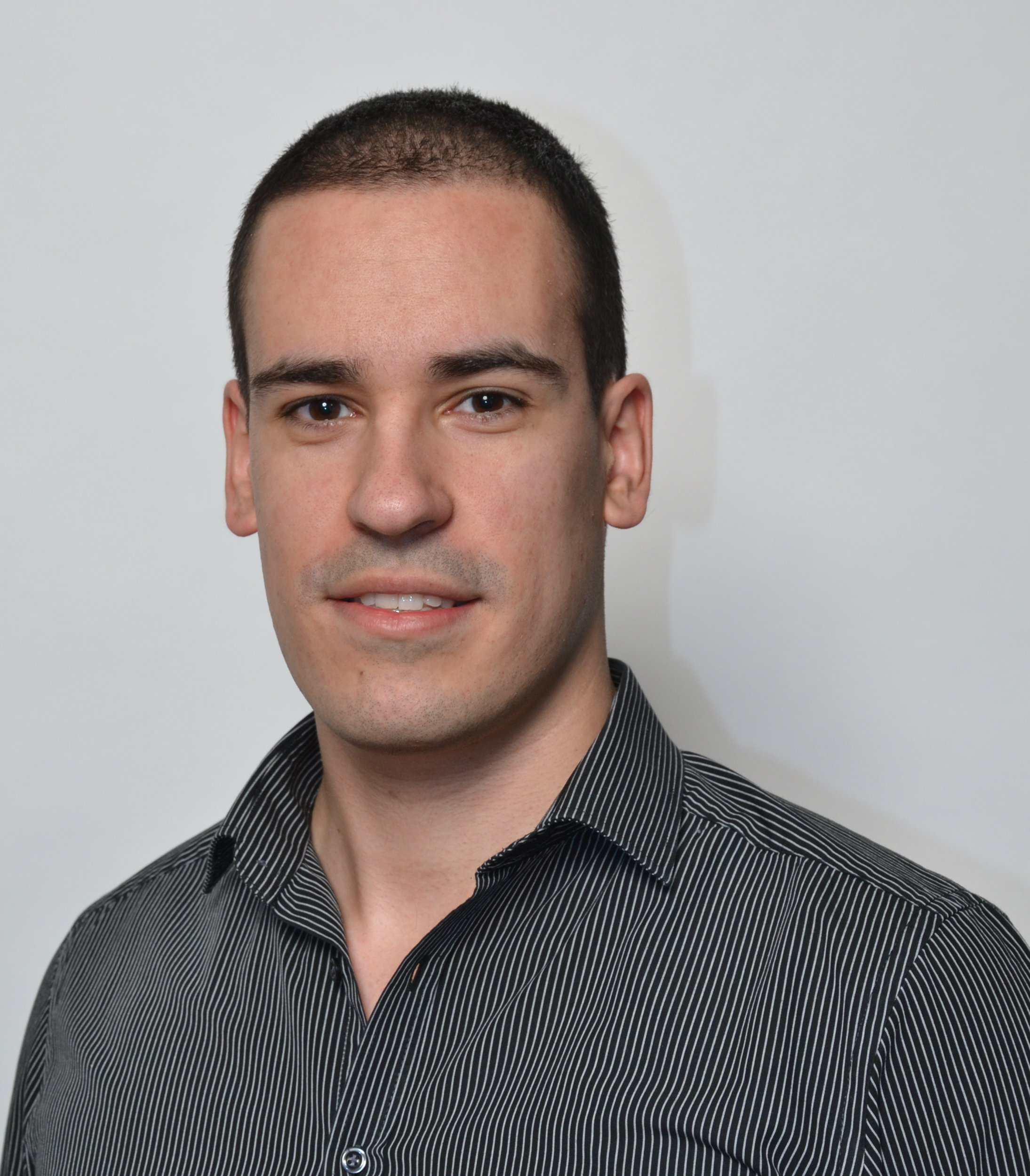}}]{Uros Markovic} (S'16) received the M.Sc. degree in Electrical Engineering and Information Technology from ETH Zurich, Zurich, Switzerland in 2016. He is currently working towards the Ph.D. degree at the Power Systems Laboratory, ETH Zurich, where he joined in March 2016. His research interests include modeling, control and optimization of inverter-based power system with low rotational inertia.
\end{IEEEbiography}
\begin{IEEEbiography}[{\includegraphics[width=1in,height=1.25in,clip,keepaspectratio]{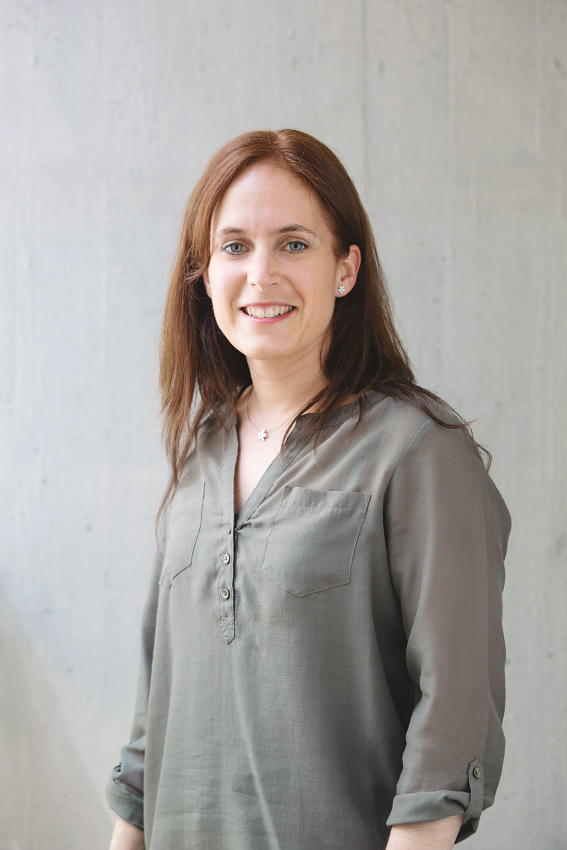}}]{Gabriela Hug} (S'05, M'08, SM'14) was born in Baden, Switzerland. She received the M.Sc. degree in electrical engineering in 2004 and the Ph.D. degree in 2008, both from Swiss Federal Institute of Technology (ETH), Zurich, Switzerland. After the Ph.D. degree, she worked in the Special Studies Group of Hydro One, Toronto, ON, Canada, and from 2009 to 2015, she was an Assistant Professor in Carnegie Mellon University, Pittsburgh, PA, USA. She is currently an Associate Professor in the Power Systems Laboratory, ETH Zurich. Her research is dedicated to control and optimization of electric power systems.
\end{IEEEbiography}
\begin{IEEEbiography}[{\includegraphics[width=1in,height=1.25in,clip,keepaspectratio]{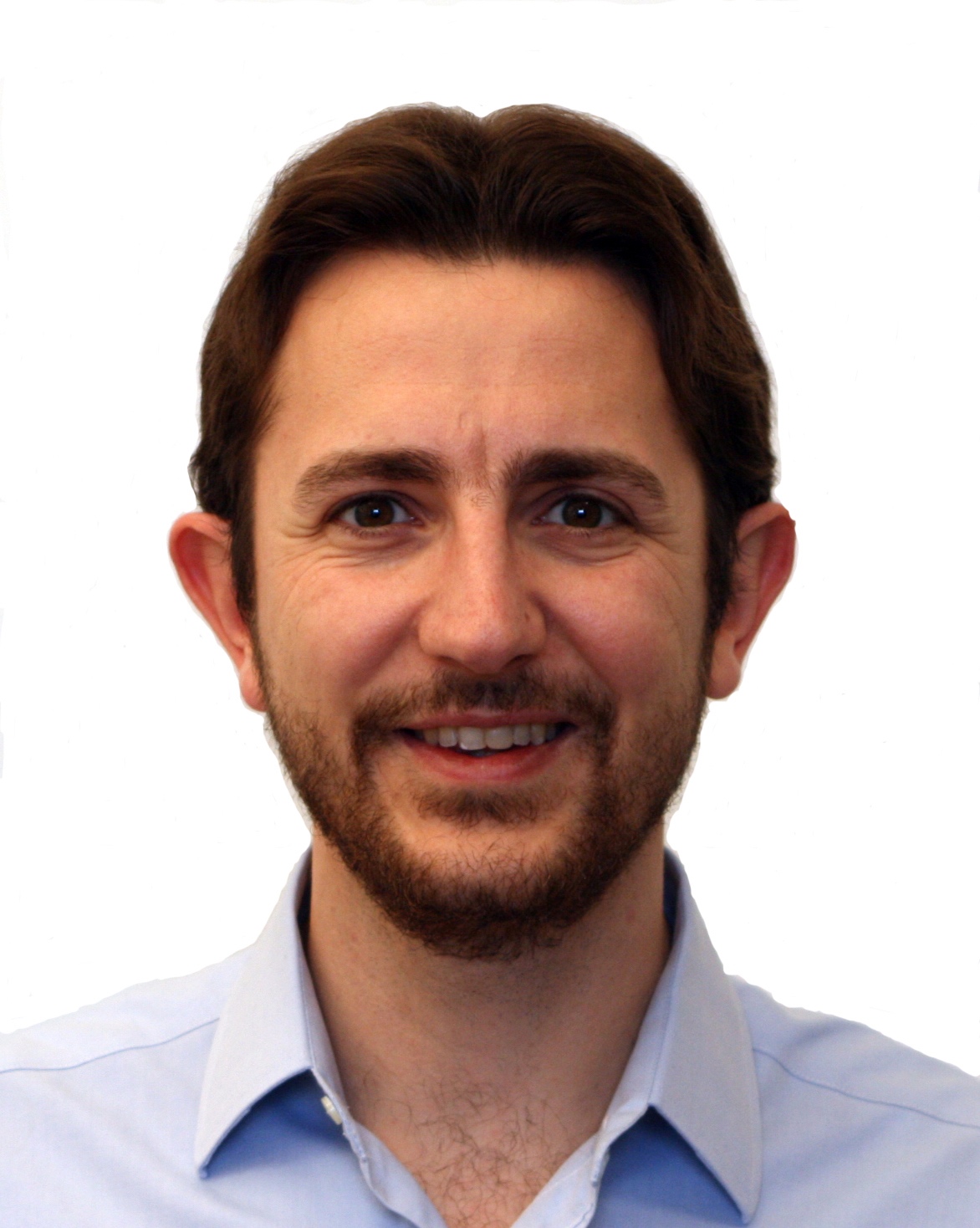}}]{Spyros Chatzivasileiadis} (S'04, M'14) is an Assistant Professor at the Technical University of Denmark (DTU). Before that he was a postdoctoral researcher at the Massachusetts Institute of Technology (MIT), USA and at Lawrence Berkeley National Laboratory, USA. Spyros holds a PhD from ETH Zurich, Switzerland (2013) and a Diploma in Electrical and Computer Engineering from the National Technical University of Athens (NTUA), Greece (2007). In March 2016 he joined the Center of Electric Power and Energy at DTU. He is currently working on power system optimization and control of AC and HVDC grids, including semidefinite relaxations, distributed optimization, and data-driven stability assessment.  
\end{IEEEbiography}

\end{document}